\documentclass[10pt]{article}

\usepackage {graphics}
\usepackage {graphicx}
\usepackage{multirow}
\newcommand{\half}{\ensuremath{{^1\!/_2}}}
\newcommand{\md}{\ensuremath{\mathrm{d}}}
\newcommand{\mb}[1]{\ensuremath{\mathbf{#1}}}
\usepackage[usenames,dvipsnames,svgnames,table]{xcolor}
 \usepackage{authblk}
  \usepackage{amsmath}

\begin{document} 

	\title{Incorporating non-adiabatic effects in Embedded Atom potentials for radiation damage cascade simulations}
	\author[1]{Daniel Mason\thanks{daniel.mason@ccfe.ac.uk}}
	\affil[1]{CCFE, Culham Centre for Fusion Energy, Abingdon, Oxfordshire OX14~3DB, United Kingdom}

	\maketitle
	
    \begin{abstract}	
    	In radiation damage cascade displacement spikes ions and electrons can reach very high temperatures and be out of thermal equilibrium.
    	Correct modelling of cascades with molecular dynamics should allow for the non-adiabatic exchange of energy between ions and electrons using a consistent model for the electronic stopping, electronic temperature rise, and thermal conduction by the electrons.    	
    	We present a scheme for correcting embedded atom potentials for these non-adiabatic properties at the level of the second-moment approximation, and parameterize for the bcc transition metals above the Debye temperature.
    	We use here the Finnis-Sinclair and Derlet-Nguyen-Manh-Dudarev potentials as models for the bonding, but the corrections derived from them can be applied to any suitable empirical potential.
    	We show with two-temperature MD simulations that computing the electronic thermal conductivity during the cascade evolution has a significant impact on the heat exchange between ions and electrons.
	\end{abstract}
	

	\section*{}
	
        A high energy ion (MeV/u) penetrating a target material loses energy by exciting electrons in the host, whereas a low energy ion (keV/u) can collide with a target atom and may start a displacement cascade. 
        As the cascade recrystallizes it leaves lattice defects which are the starting point for long-term microstructural evolution of the material. 
        The penetration range of the ion depends on the electronic stopping power, and the degree of recrystallization is strongly dependent on the cooling rate, which in turn depends on how hot ions lose energy to electrons.
  		Incorporating non-adiabatic effects into a Molecular Dynamics (MD) simulation requires the consideration of four related issues: electronic stopping (including electron-phonon coupling), electronic heat capacity, spontaneous phonon emission(heat return from electrons to ions), and electronic thermal conductivity.
  		
  		Following earlier MD simulations coupling atoms to a constant temperature electronic subsystem by Caro \& Victoria \cite{CaroVictoria_pra1989}, and Finnis et al\cite{PhysRevB.44.567}, a full two-temperature MD model was introduced by Ivanov and Zhigilei\cite{Ivanov_PRB2003} to model femtosecond laser pulse melting. 
        Electronic stopping and heat return were handled by a single friction coefficient proportional to the temperature difference between ions and electrons in a manner analogous to Finnis et al.\cite{PhysRevB.44.567}, with electronic heat capacity linear in electronic temperature and thermal conductivity proportional to the quotient of electronic and ionic temperatures, equivalent to the reasonable assumption that electron-phonon scattering dominates conductivity in their experiment.
        
        Later Duffy \& Rutherford\cite{Duffy_jpcm2007,Rutherford_jpcm2007} modelled heat-exchange between ions and electrons with a friction term on the ions and stochastic return.
        The friction was taken to be small for low kinetic energy ions- representing electron-phonon coupling, and larger for high energy ions- representing electronic stopping.
        The separation between these regimes was (somewhat arbitrarily) set to twice the cohesive energy, following Zhurkin \& Kolesnikov\cite{Zhurkin_NIMB2003}.
        For the electronic specific heat capacity a tanh function was employed, $C_e = 3k_B \tanh( 2\times10^{-4}\, T_e )$, which shows a metallic linear regime saturating at high temperature.
        Heat return from electrons to ions was modelled with an inhomogeneous Langevin thermostat, and thermal conductivity was taken to be a constant.
        They were able to demonstrate for model iron simulations that the number of residual defects remaining after the cooling of a 10keV displacement cascade increased with ionic cooling due to friction and decreased with the annealing due to warm electrons.        
		Phillips \& Crozier\cite{Phillips_JCP2009,Phillips_JCP2010} subsequently corrected an energy drift in Duffy \& Rutherford's original implementation and integrated 2TMD into LAMMPS\cite{LAMMPS}. 
        
		Zarkadoula et al\cite{Zarkadoula_JPCM2013,Zarkadoula_JPCM2014} tweaked the physics in the Duffy \& Rutherford model by replacing the constant thermal conductivity with one which scales with the electronic heat capacity- equivalent to assuming a constant electronic scattering time through the simulation. 
		Comparing the full 2TMD model to a simpler `friction' model, where energy is merely removed from high energy ions ( above twice cohesive energy ) and not returned, they showed that the lower energy electron-phonon coupling reduced both the size of the maximum damage region and the time taken to return to thermal equilbrium.
		This resulted in a similar total number of Frenkel pairs for each model but with different clustering of vacancies and interstitials.

         While these, and other, simulations\cite{0965-0393-6-5-003,PhysRevB.50.13194,PhysRevB.57.R13965,Duvenbeck_njp2007,Sand_EPL2013,lePage2009} show that non-adiabatic effects might be significant, there is little self-consistency in how the stopping relates to heat capacity and heat capacity relates to conductivity.
 		This paper seeks to tackle electronic stopping, temperature rise, and thermal conductivity in a single consistent framework suitable for use in two-temperature Molecular Dynamics.
 		We parameterize for the popular Ackland-Thetford (AT)\cite{Ackland_PMA1987} and Derlet-Nguyen-Manh-Dudarev (DND) potentials\cite{Derlet_PRB2007} for bcc transition metals.

		A two-temperature model simulating radiation damage in metals should incorporate the following physics:
		\begin{description}
			\item {Electronic stopping}\newline
				At low temperatures light electrons move quickly compared to heavy ions, and it is a good approximation to assume the electrons will readily find their ground state.
				This assumption is called the Born-Oppenheimer approximation.
				However it can not be exactly the case if the atoms are moving, as the ionic potential changes the electrons can never be instantaneously in the ground state. 
				Thus they must be in a somewhat higher energy state- energy is transferred constantly from ions to electrons. This process is known as electronic stopping. 
				Some authors distinguish two regimes- electron-phonon stopping where the ions have low energy and so are vibrating about lattice sites and electronic stopping where ions move so fast the lattice is (to them) irrelevant and they can be treated as moving in a Fermi gas.
				The physical principle is the same in either case, only the simplifying approximations change \cite{Race2010b}.
				We derive a formula for electronic stopping in both regimes in section \ref{stopping_section}.
				
			\item {Electronic heat capacity}\newline
				Energy is transferred from ions to electrons by exciting electrons from occupied states below the Fermi surface to unoccupied states above the Fermi surface.
				Fermi's Golden Rule is illuminating here- if ions are oscillating with a phonon frequency $\Omega$ then (in the limit of long time) an electronic transition shall be excited across an energy gap $\hbar \Omega$.
				Therefore if ions are moving fast a spectrum of high frequency transitions may be stimulated, which promotes electronic transitions from a wide range of the local electronic density of states\cite{Mason_NJP2012}.
				For general keV to MeV ions there is no \emph{a priori} reason to assume the spectrum of excitations forms any easily recognisable pattern.
				However if the bulk of ions are merely at high temperature, the electronic transitions are many and small.
				Then the electronic excitation takes the form of a diffusion of the occupation of states in energy space\cite{Race2009}.
				This diffusion looks very similar to a thermal spectrum of occupations, so while it is important to remember the electrons are not in thermal equillibrium, there is an identifiable pseudotemperature associated with the electronic excitation typical from high temperature ions\cite{Lin2009}.
				
				If the heat capacity of the electrons is estimated using the Sommerfeld expansion as $C_e \equiv \partial \epsilon/\partial T_e =  {}^{\pi^2}\!\!/\!_{3} k_B^2 D(\epsilon_F) T_e$, then we can associate a (true) temperature rise with a given energy transfer.
				The electronic pseudotemperature generated by energy transfer from high energy ions is found to match the required true temperature rise \cite{Race2009}.
				Therefore it is quite sensible to use electronic pseudotemperature and heat capacity as simple proxies for the complex pattern of electronic excitations which occur in radiation damage.
				Heat capacity can be readily computed from the local density of states in section \ref{heatCap_section}.
				It should be highlighted that this electronic energy should manifest as a change in the bonding of the atoms- a high temperature corresponding to a reduction in the attractive energy.
				
			\item {Spontaneous phonon emission}\newline
				The process of electronic stopping via stimulated electronic excitation and its reverse process, stimulated phonon emission, are captured when classical ions move in the mean field potential of quantum mechanical ions, and accounts for the majority of the energy transfer between hot ions and cold electrons \cite{lePage2008}.
				However quantum mechanical ions permit an additional important mode of energy transfer, namely spontaneous phonon emission, and this is necessary to produce thermal equilibrium between ions and electrons.
				The rate of this process must depend on the electron density of states, the phonon density of states and the coupling matrix elements between them.
				
				We expect tunnelling and zero-point energy effects to be significant where light elements are involved, but here we make no attempt to include quantum mechanical phonons explicitly into molecular dynamics.
				The reader is referred to refs \cite{Horsfield_jpcm2005,McEniry2010,Stella_PRB2014,Heatwole_JPSJ2008}.
				
				We instead derive in section \ref{SPE} the Fokker-Plank equation for the ion-electron energy transfer assuming a Gaussian white noise return, and so treat spontaneous phonon emission using an equation entirely analogous to that used in dissipative particle dynamics.
				
			\item {Electronic thermal conductivity}\newline
				Thermal conductivity in metals is dominated by electrons, and so it is the electrons which are responsible for taking heat out of a radiation damage cascade not the phonons\cite{Duvenbeck_prb2005,Duvenbeck_nimb2007}.
				The quantum mechanics of this process is extremely complicated\cite{Michel_prl2005,Dubi_pre2009}, but the practical upshot is well-known as Fourier's law.
				The local thermal conductivity is proportional to both the local heat capacity and the electron scattering time.
				The electron scattering rate is the sum of the electron-electron scattering rate, the electron-phonon scattering rate and the electron-impurity scattering rate.
				Electron-electron scattering is an electronic quasiparticle effect beyond the scope of this paper- we fit a simple functional form to the experimental thermal conductivity.
				Electron-phonon scattering however must be related to the electronic stopping, so a consistent environmentally dependent formula is proposed in section \ref{eph_scattering}	.
				Electron-impurity scattering is difficult to model but is the major determining factor of thermal conductivity in disordered systems.				
				A detailed model is beyond the scope of this paper, but we offer a very simple model for impurity scattering as a placeholder in section \ref{eimp_scattering}.
				
		\end{description}

	\section{Electronic Stopping}
		\label{stopping_section}
		
		Our goal is to produce simple formulae for adding the effect of heat transfer between electrons and ions which can be added to existing empirical potentials used in molecular dynamics simulation.
		It is therefore not appropriate to start with a full quantum mechanical picture of ions and electrons together, but rather we start with classical ions and quantum mechanical electrons, the so-called Ehrenfest approximation\cite{Todorov2001}.
		The Ehrenfest approximation accurately reproduces the true electron-ion interaction when ion temperatures are high, as is the case in a radiation damage cascade\cite{lePage2008}.
		We treat the purely quantum mechanical process of electron-to-ion energy transfer via spontaneous phonon emission later.

        The embedded atom potential form expresses the energy of the system of atoms as	
			\begin{equation}
				E\left( \mb{R} ; t \right)_{\mathrm{EAM}} = \half \sum_i M_i \left| \dot{\mb{R}}_i \right|^2 + V\left( \mb{R} \right) + \sum_i F\left[ \rho_i \right],
			\end{equation}
		where $\rho_i = \sum_{j \in \mathcal{N}_i} \phi( \mb{R}_{ij})$ is a scalar proportional to the electron density into which atom $i$ is embedded.
		For the original Finnis-Sinclair form\cite{Finnis1984} the embedding function is a simple square root function $F\left[ \rho \right] = -A \sqrt{ \rho }$.
		If we assume a Freidel rectangular d-band model with $N_e$ electrons in $N_a$ states, width $W_i$ and height $D_i = N_a/W_i$ with centre of band $\alpha_i$,
		the zero-temperature bond energy is given by
			\begin{equation}
				\label{zeroTempCohesiveBondEnergy}
				F_{T=0,i} = -A \sqrt{ \rho_i } = \int_{\alpha_i-W_i/2}^{\lambda} 2 D_i(\epsilon) (\epsilon-\alpha_i)  \md \epsilon = - \frac{N_e W_i}{2} + \frac{N_e^2 W_i}{4 N_a} 
			\end{equation}
		hence the width of the d-band is given by 
			\begin{equation}
				W(\rho) = \frac{ 4 A N_a }{ N_e (2 N_a - N_e ) } \sqrt{\rho} \equiv w \sqrt{\rho}, 
			\end{equation}
		defining the constant $w$.		
		
		In \ref{StoppingCalc} we derive the electron-phonon damping force using a simple tight-binding model with explicit electronic evolution, then transfer the result at the end of the section to this EAM potential.
		The electron-phonon damping force is shown to be
        \begin{equation}
                \label{lagForce}
                \mb{F}_{\mathrm{e-ph},i} 
                    = \frac{  \bar{W} }{W_i} \sum_{j \in \mathcal{N}_i} \, \mb{B}_{ij} ( \dot{\mb{R}}_j-\dot{\mb{R}}_i )
            \end{equation}		
		
        At low temperatures the damping tensor is 
            \begin{equation}
                \lim_{k_B T_e \ll W_i} \mb{B}_{ij} =  \zeta \frac{ 8.4719 N_a w^2 \hbar  }{W_i W_j} \nabla_i \phi(\mb{R}_{ij}) \nabla_i \phi(\mb{R}_{ij})^T.
            \end{equation}

		We can fit the empirical constant $\zeta$ to experimental data as follows.
		The rate of heat transfer between ions and electrons ( excluding source terms ) is described by coupled differential equations
			\begin{equation}
			    \label{heatTransferEqn}
				C_{^{e}\!/\!_I} \frac{\partial T_{^{e}\!/\!_I}}{\partial t} = \kappa_{^{e}\!/\!_I} \nabla^2 T_{^{e}\!/\!_I}  \pm G \left( T_I - T_e \right),
			\end{equation}
		where $G$ is the electron-phonon coupling factor.
		
		From equation \ref{lagForce} we have for atom $i$ in thermal equilibrium		
			\begin{equation}
				\left. \left< \frac{\partial E_e}{\partial t} \right> \right|_{e-ph} 
				    = \left< \frac{  \bar{W} }{W_i} \sum_{j \in \mathcal{N}_i}  \, \mb{B}_{ij} ( \dot{\mb{R}}_j-\dot{\mb{R}}_i ) \cdot \dot{\mb{R}}_i \right>  \equiv \Omega G_{\mathrm{EAM}} \,  T_{I},
			\end{equation}    		
		We can evaluate this expression where the atoms are on their ideal lattice sites, assuming $\left< \dot{\mb{R}}_j \cdot \dot{\mb{R}}_i \right>=\delta_{ij}\, 3k_B T_I/M_I$, giving
		    \begin{equation}
				\label{GEAM}
		        G_{\mathrm{EAM}}  = \frac{ 1 }{\Omega} \bar{B} \frac{ 3 k_B }{M_I},
		    \end{equation}
		where we have taken the expected magnitude of the damping tensor ($\bar{B}$) to be one-third its trace at the perfect zero-temperature lattice:
			\begin{equation}
				\bar{B} = \left. \frac{1}{3} \left< \mathrm{Tr}\left[ \sum_{j \in \mathcal{N}_i} \, \mb{B}_{ij} \right] \right> \right|_{\mathrm{perfect}}. 
			\end{equation}

		A low temperature estimate for the electron-phonon coupling factor per unit volume is given by Allen\cite{Allen_PRB1987,Allen_PRL1987}:
			\begin{equation}
				\label{AllenEPhFactor}
				G_A = \frac{\pi \hbar k_B}{\Omega} \lambda \langle \omega^2 \rangle \, D(\epsilon_F),
			\end{equation}			
		where $\lambda$ is the first reciprocal moment of the Eliashberg spectral function and $\langle \omega^2 \rangle$ the second moment of the phonon density of states.
		These factors may be measured experimentally or computed using DFT\cite{Daraszewicz_APL2014}.
		Equating $G_{\mathrm{EAM}} = G_A$ gives the empirical fitting parameter $\zeta$
			\begin{equation}
				\zeta = \frac{ \pi \lambda \langle \omega^2 \rangle 2 M_I  \bar{W}}
							{  8.4719 w^2 \left< \mathrm{Tr}\left[  \nabla_i \phi(\mb{R}_{ij}) \nabla_i \phi(\mb{R}_{ij})^T \right] \right>  }.
			\end{equation}
		Values for $\zeta$ are given in table \ref{electronicStopping}, completing a parameterization for electron-phonon stopping from the form of the EAM potential.
        Note that by construction $\bar{B}$ is material-dependent but not potential-dependent.	The environmental dependence of $\mb{B}$ \emph{is} potential-dependent.
					
        The value for $\bar{B}$ in iron corresponds to a coupling strength $\chi =1.2$ps$^{-1}$ in ref\cite{Rutherford_jpcm2007}, which gives the peak in the expected number of residual defects per cascade.

	\begin{table}
        \centering
        \small
        \begin{tabular}{l|lll|lll|lll}
        	Element	&	\multicolumn{3}{c|}{experiment} 					&	\multicolumn{3}{c|}{e-ph regime}        & \multicolumn{3}{c}{electronic stopping regime}  			\\
        			&	$\lambda$	&			$\lambda\langle \omega^2\rangle$& $G_A$				&$\zeta_{AT}$&$\zeta_{DND}$	& $\bar{B}$ 				 & $\delta e_c$	&	$k_c$	&	$\tilde{B}$		\\
        			&				&	meV$^2$							& eV/K/fs/\AA$^3$	&			&		        &  eV fs/\AA$^2$ 					  & eV			&	eV		&	eV fs/\AA$^2$	        \\
        	\hline      							    				            	                        						            			        	
        	V       &	0.8 	&	281.6 	a)						&	1.49E-08		&	0.094   &	  0.903     &  4.281				     & 0.25			&	70		&	8.70 	\\
        	Nb      &	0.88	&	275   	b)						&	1.09E-08		&	0.086   &	  0.860     &  5.524     				& 0.30			&	200		&	8.46 	\\
        	Ta      &	0.88	&	165.44	c)						&	4.93E-09		&	0.124   &	  1.949     &  6.466     				& 0.35			&	550		&	12.84	\\
        	Cr      &	0.13	&	128.31	a)						&	3.00E-09		&	0.549   &	            &  0.751     				& 0.33			&	100		&	6.22 	\\
        	Mo      &	0.44	&	118.8 	d)						&	1.88E-09		&	0.048   &	  0.998     &  1.151     				& 0.67			&	990		&	8.45 	\\
        	W       &	0.26	&	110.5 	a)						&	1.25E-09		&	0.047   &	  1.496     &  1.188     				& 1.00			&	4200	&	21.16	\\
        	Fe		&	0.9 	&	738    e)						&	2.60E-08		&	1.990   &	            &  6.875     				& 0.50			&	260		&	5.88 	\\
        \end{tabular}                                                                                                                                                                                                          
        \caption{                
        	Electronic stopping parameters for the materials considered. 
        	Experimental parameters from a) \cite{Brorson_prl1990} b) \cite{Fletcher_prb1988} c) \cite{AlLehaibi_prb1987} d)\cite{Mazin_JphysF1984} e)\cite{Allen_PRB1987}.
        	The values for $\tilde{B}$ are extracted from the low velocity self-ion stopping limit of the SRIM code\cite{Ziegler:1985fj}.     	        	
        	The kinetic energy for transition between electron-phonon and electron-stopping regimes ($E_c$) is given at $T_e=0$.
        }
        \label{electronicStopping}
    \end{table}        
        
	\subsection{Electronic stopping of high energy/high temperature ions}
		\label{estopping}
        			
		For very high energy ions, which excite larger electronic transitions far from the Fermi surface, the electron-phonon coupling in equation \ref{e-ph_stopping} may not reproduce observed electronic stopping.
		This may seem bad news for a radiation damage simulation where the initial ions have very large kinetic energies in the keV range, but there are few such atoms, and their lifetime is short. 
		It is the large number of lower energy athermal ions in the 1-10eV range which are important in determining the heat transfer\cite{Duvenbeck_prb2005,Duvenbeck_nimb2007}.
		The SRIM code\cite{Ziegler:1985fj} is well-known to accurately predict the range and straggling of high energy ions in matter, and it does this with a model for stopping appropriate for ions with sufficiently high energy that the directional bonding in the lattice may be replaced with a isotropic electron density\cite{Firsov:1959nh,LindhardScharff_pr1961,CaroVictoria_pra1989}. In this limit
			\begin{equation}
				\mb{F}_{\mathrm{e-stopping},i} = - \tilde{B} \dot{\mb{R}}_i	
			\end{equation}

\begin{figure}[h!t!b]
           \begin {center}
               \includegraphics [width=5.0in] {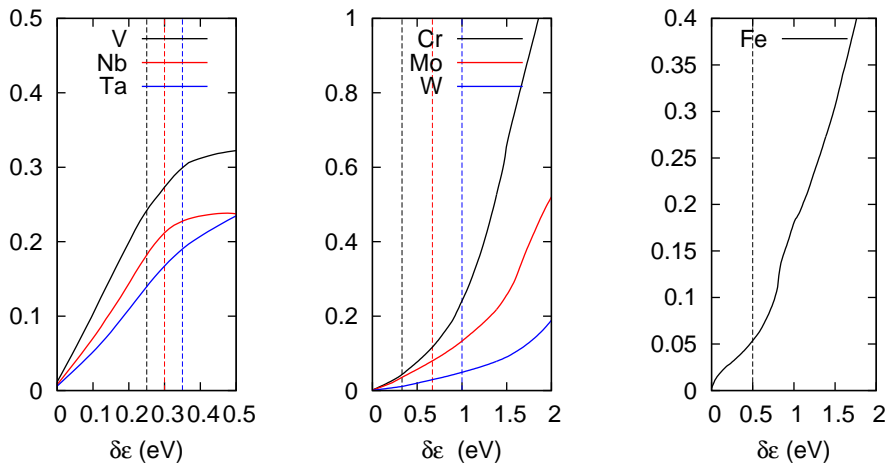}
           \end{center}
           \caption {
           Electronic joint densities of states for the transition metals considered, with the extent of the linear region marked by a dotted vertical line.
            }
           \label {jdos}
       \end {figure}

		Finding the exact point at which the electron-phonon model fails is complicated, but the electronic joint density of states $J(\delta \epsilon)$ gives part of the answer.
			\begin{equation}
				J(\delta \epsilon)	= \int_{\epsilon=-\infty}^{\epsilon_F} \md \epsilon \, D(\epsilon) D(\epsilon+\delta e) \Theta( \epsilon_F - \epsilon - \delta \epsilon ),
			\end{equation}
		where $\Theta(x)$ is the Heaviside function, $\Theta(x<0) = 1,\Theta(x>0) = 0$. This is plotted for the perfect lattice in figure \ref{jdos}.
		As $J(\delta \epsilon)$ is a measure of the number of accessible electronic transitions across $\delta \epsilon$, we can use it as a zeroth order approximation to the electronic stopping power.
		An electronic transition $\delta \epsilon$ will be excited by an atom travelling at velocity $v$ where $\delta \epsilon = h v/d$, and $d=\sqrt{3/4}a_0$ is the distance between atoms in a bcc lattice.
		$J(\delta \epsilon)$ is approximately linear for small transition energies $\delta \epsilon$, indicating a linear increase in the number of accessible transitions with $\delta \epsilon$, and so in this region electronic stopping power is proportional to $v$.
	   	Past some cutoff $\delta \epsilon_c$ the linearity fails, suggesting that band effects must be important in this region.
	   	This band cutoff corresponds to an ionic kinetic energy at which to transfer from electron-phonon to electronic stopping models
	   		\begin{equation}
	   		    \label{eph_limit}
	   			k_{c}(T_e) = \frac{3}{8} M \left( \frac{ a_0 \delta \epsilon_c }{h} \right)^2.
	   		\end{equation}
	   	This predicts a transition regime for ions with kinetic energies of hundreds of eV (see table \ref{electronicStopping}).
	   	There is no support for a kinetic energy cutoff of the order twice the cohesive energy.
	   	   	
	   	A second important energy scale is where an atom can excite electrons across the whole band- ie where $\delta \epsilon = h v/d = \bar{W}$, a kinetic energy of order
	   		\begin{equation}
	   		    \label{band_limit}
	   			k_{c,band}(T_e) = \frac{3}{8} M  \left( \frac{ a_0 \bar{W} }{h} \right)^2.
	   		\end{equation}
    	This predicts a model failure point for ions with much higher kinetic energies- ranging from 40keV for Fe to 2.3MeV for W.
	   		
	   	As the increase in stopping in the SRIM model is due to sampling a greater range of the bandwidth, we should also transition from electron-phonon damping to frictional damping at high temperature where $k_B T_e \sim \delta \epsilon_c$. 
	   	
        We should not expect to see any fine-structure in stopping power, as at higher energies the average stopping from many accessible transitions is felt.
   		We therefore recommend avoiding finite-band-width effects whether caused by high temperature or swift ions by smoothly transitioning the stopping from electron-phonon regime to electronic stopping regime as:
			\begin{eqnarray}
				\mb{F}_{e-ph,i} &=& \frac{  \bar{W} }{W_i} \sum_{j \in \mathcal{N}_i} \, \mb{B}_{ij} ( \dot{\mb{R}}_j-\dot{\mb{R}}_i ) \min\left( p\left( 2 - \frac{E_k}{k_c} \right),f\left(-\delta \epsilon_c;T_e\right)\right)		\nonumber	\\
								&& - \tilde{B} \dot{\mb{R}}_i \left(1-\min\left( p\left( 2 - \frac{E_k}{k_c} \right),f\left(-\delta \epsilon_c;T_e\right)\right)\right),
			\end{eqnarray}
		where $E_k=\half M \left| \dot{\mb{R}}_i \right|^2$ is the kinetic energy, $f(\epsilon;T_e)$ is the Fermi-Dirac function, and 
			\begin{equation}
				p(x) = \left\{ 	\begin{array}{cc} 
								0				&		x<=0	\\
								3x^2-2x^3		&		0<=x<=1	\\
								1				&		x>=1	
								\end{array}	\right.
			\end{equation}
        This smooth transition also alleviates any possible issues arising from the decorrelation of the coupling between occupied and unoccupied states in equation \ref{decorrelation_eqn}.
						
	   	As indicated above, this is only part of the answer, as the coupling between states is not taken into account in this simple picture, nor are the defect energy levels introduced by disorder in the crystal.
        It should also be remembered that swift ions in this model are strictly charge neutral, which may not be the case in reality.
	   	
	    For a more accurate description, time-dependent DFT calculations can be employed to find the electronic stopping power at high and low velocity\cite{Pruneda_prl2007,Campillo_prb1998,AhsanZeb_prl2012}.

\begin{figure}[h!t!b]
           \begin {center}
               \includegraphics [width=4.5in] {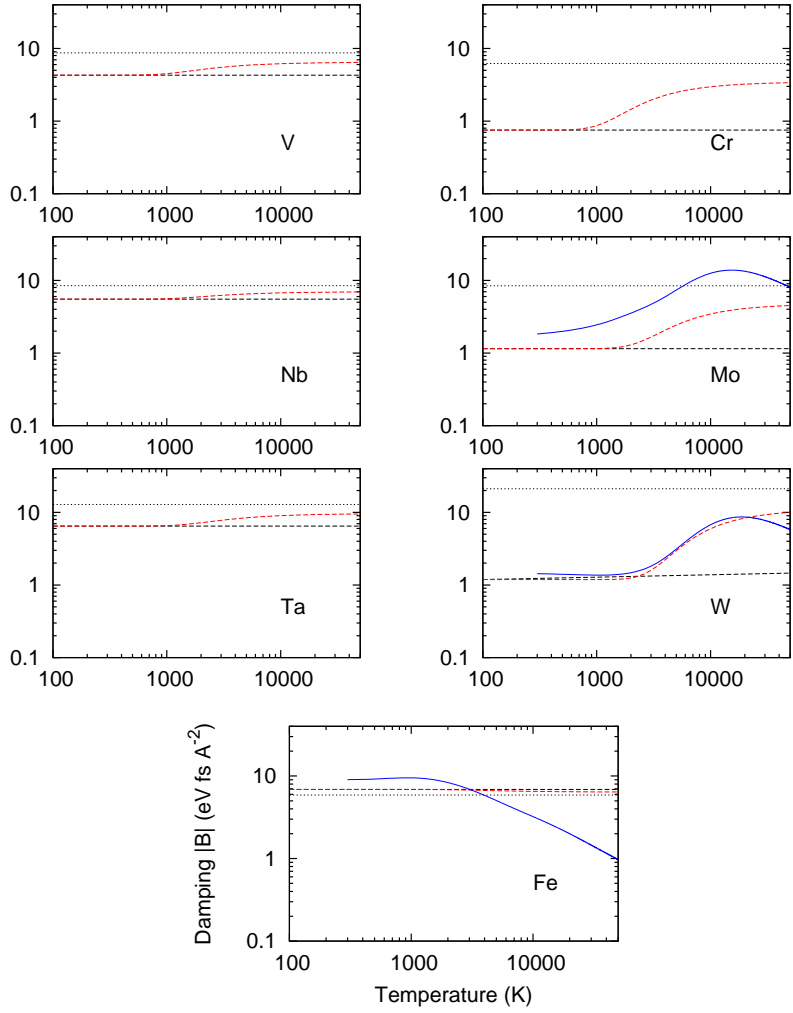}
           \end{center}
           \caption {
           Electronic stopping as a function of temperature in $K$, defined as the constant $B$ where $\mathbf{F}_i \sim - B \dot{\mb{R}}_i$, in units $eV/\AA$.
           The lattice is assumed to be perfect, with ions and electrons in thermal equilbrium.
           Two horizontal lines correspond to the low temperature value extracted from equation \ref{AllenEPhFactor}, and the SRIM value in table \ref{electronicStopping}.
           The SRIM value is the higher in all cases except iron.
           The solid blue lines are the DFT estimates of Lin\cite{Lin_PRB2008}. It should be noted that Lin computes the result for spin-degenerate iron.
           Lines for EAM AT and DND parameterizations are almost indistinguishable.           
            }
           \label {stopping}
       \end {figure}

	   	The expected stopping as a function of temperature is plotted in figure \ref{stopping}.

	\section{Electronic Heat Capacity}
		\label{heatCap_section}
	
		We now turn our attention to finding the dependence of the potential energy on the electronic pseudotemperature.
        Temperature dependence in the second-moment model was introduced by Khakshouri et al\cite{Khakshouri_prb2008}, who used an explicitly rectangular d-band local density of states of height $D_i^{(d)} =  N_a/W_i$ between $\epsilon=\alpha_i - W_i/2$ and $\epsilon=\alpha_i + W_i/2$.
        They then added a s-p band from $\epsilon=\alpha_i + W_i/2$ to $\epsilon=\infty$ with the same height $D_i^{(s-p)} = N_a/W_i$. This simplification allows for exact solution of the integrals.
        Here we rederive the equations, correcting an error in ref\cite{Khakshouri_prb2008}, and examining the limits in detail.
        
        The number of electron orbitals would be $N_a = 5$ for a d-band, although we shall see later it is useful to keep this as a parameter of the model to fit the density of states.
        The number of electrons $N_e$ fixes the difference between the chemical potential $\lambda$ and the bottom of the band, as
            \begin{equation}
                N_e = 2 \frac{N_a}{W_i} \int_{\alpha_i - ^{W_i}\!\!/\!_2}^{\infty} f(\epsilon;\lambda,T_e) \md \epsilon,
            \end{equation}
        requires
            \begin{equation}
                \frac{\lambda - (\alpha_i - W_i/2)}{k_B T_e} = \log \left[ \exp \left( \frac{N_e W_i}{2 N_a k_B T_e} \right) - 1 \right] \equiv \mu_i,
            \end{equation}
        defining the useful ratio $\mu_i$.
        
        However this ratio is hiding a physical inconsistency, namely that the electronic pseudotemperature can apparently be high while the bandwidth tends to zero, which may b e the case if an ion is ejected from a surface at non-zero temperature. 
        In the original Finnis-Sinclair formalism, there is no problem with ejecting an ion: the d-band becomes a delta-function centred on the Fermi level as the atom leaves the influence of its neighbours, and the ion energy tends to zero.
        In the Khaksouri formalism, the d-band also tends to a delta function, but the assumed s-p band attached at the top of the band, included to make for closed integral forms, does not: it is always infinite in width but now becomes infinite in height also.
        This requires the electrons to be in the flat high-energy tail of the Fermi-Dirac function, and therefore tending to infinite entropy, and -infinite free energy.
       
        To resolve this, we can make the \emph{ad hoc} requirement that the temperature can not be much greater than the bandwidth, when the bandwidth is small.		        							        		
        	\begin{eqnarray}
        		\label{temperatureRestriction}
        		k_B T_{e,i} &\le& y \equiv \frac{ W_i }{ W' }\quad, \mathrm{if}\quad W_i<\half \bar{W}	\nonumber		\\
        				& = & 	\left\{ 	\begin{array}{lr}	\displaystyle		
        								k_B T_e		&	k_B T_e \le \frac{y}{2}		\\
        								y \sum_{j=0}^4 a_j \left( \frac{ k_B T_e } { y } \right)^j	&	\frac{y}{2} \le k_B T_e \le \frac{3 y}{2} 	\\
        								y	&	k_B T_e \ge \frac{3 y}{2} 
        									\end{array}
        						\right.	\nonumber	\\
        	\mbox{with}	\quad\quad
					a_j & = & \frac{1}{32} \left\{ 5,0,72,-64,16 \right\}.
        	\end{eqnarray}
        Then $\mu_i$ has the limits
            \begin{eqnarray}
                \lim_{T_e/W_i\rightarrow 0} \mu_i &=& \left( \frac{N_e W_i}{2 N_a k_B T_e} \right) \rightarrow +\infty \nonumber   \\
                \lim_{W_i/T_e\rightarrow 0}  \mu_i &\equiv& \lim_{ W_i/T_e \rightarrow W' } \mu_i = \mu_0.
            \end{eqnarray}
		Making such an \emph{ad hoc} stipulation is somewhat unsatisfactory, but in practice it will not be invoked unless atoms are to co-exist in the gas phase.            
		We include it here as a mechanism for completing a self-consistent, well-behaved formalism for safe MD simulation.
        A better model for temperature dependence in the potential energy would build in smooth limits, and may include the possibility of ejected atoms becoming ionised. 
        However this is beyond the scope of the current paper.
        
        Values for $W'$ can be found by considering the free energy, which is done in \ref{electronicEntropy}, and the results given in table \ref{parameterization}.

        The cohesive bond energy for atom $i$ is then\cite{Sutton_jpc1988} 
            \begin{eqnarray}
                F_{T_e,i} &=& 2 \int_{-\infty}^{\infty} D(\epsilon) f(\epsilon;\lambda,T_{e}) (\epsilon-\alpha_i) \md \epsilon          \nonumber   \\
                    &=& 2 \frac{N_a}{W_i} \int_{\alpha_i - ^{W_i}\!\!/\!_2}^{\infty} f(\epsilon;\lambda,T_{e}) (\epsilon-\alpha_i) \md \epsilon          \nonumber   \\
                  &=& - \frac{ N_e W_i }{2}  + 2 \frac{N_a (k_B T_{e,i})^2}{W_i}  \left[ \frac{\pi^2}{6} + \frac{ \mu_i^2 }{2} + \mathrm{Li}_2\left( -e^{-\mu_i} \right) \right],
            \end{eqnarray}
        where $\mathrm{Li}_2$ is the dilogarithm function
            \begin{equation}
                \mathrm{Li}_2 \left( -e^{\mu} \right) = -\int_0^{\infty} \frac{x}{e^{x-\mu} + 1} \md x,
            \end{equation}
        which has the limiting behaviour
            \begin{equation}
                \lim_{T_e/W_i\rightarrow 0 , \mu\rightarrow\infty} \mathrm{Li}_2 \left( -e^{-\mu} \right) =  -e^{-\mu} = 0                 
            \end{equation}

		We can write this as a thermal correction to the original zero-temperature energy
			  \begin{eqnarray}
                	F_{T_e,i} &=& F_{0,i} + \Theta_{ T_{e},i }				\nonumber	\\
               \mbox{where} \quad
               	\label{temperatureDependentPotential}
               		\Theta_{ T_{e},i } &=& -\frac{N_e^2 W_i}{4 N_a} + \left(\frac{2 N_a}{W_i}\right) (k_B T_{e})^2 \left[ \frac{\pi^2}{6} + \frac{\mu_i^2}{2} + \mathrm{Li}_2\left( -e^{-\mu_i} \right) \right],
              \end{eqnarray}
        with $F_{0,i}$ being the zero temperature cohesive bond energy linear in $W_i$ given by equation \ref{zeroTempCohesiveBondEnergy}.            
        At low and high temperatures the correction is:
            \begin{eqnarray}
                \label{lowandhighTenergy}
                \lim_{T_e/W_i\rightarrow 0} \Theta_{ T_{e},i } &=&  \frac{\pi^2 (k_B T_e)^2}{6}  \left(\frac{2 N_a}{W_i}\right) \nonumber \\
                \\
                \lim_{W_i/T_e \rightarrow 0} \Theta_{ T_{e},i } &=&  -\frac{N_e^2 W_i}{4 N_a} + \left(\frac{2 N_a}{W'}\right) (k_B T_{e}) \left[ \frac{\pi^2}{6} + \frac{\mu_0^2}{2} + \mathrm{Li}_2\left( -e^{-\mu_0} \right) \right]
            \end{eqnarray}
           
\begin{figure}[h!t!b]
           \begin {center}
               \includegraphics [width=4.8in] {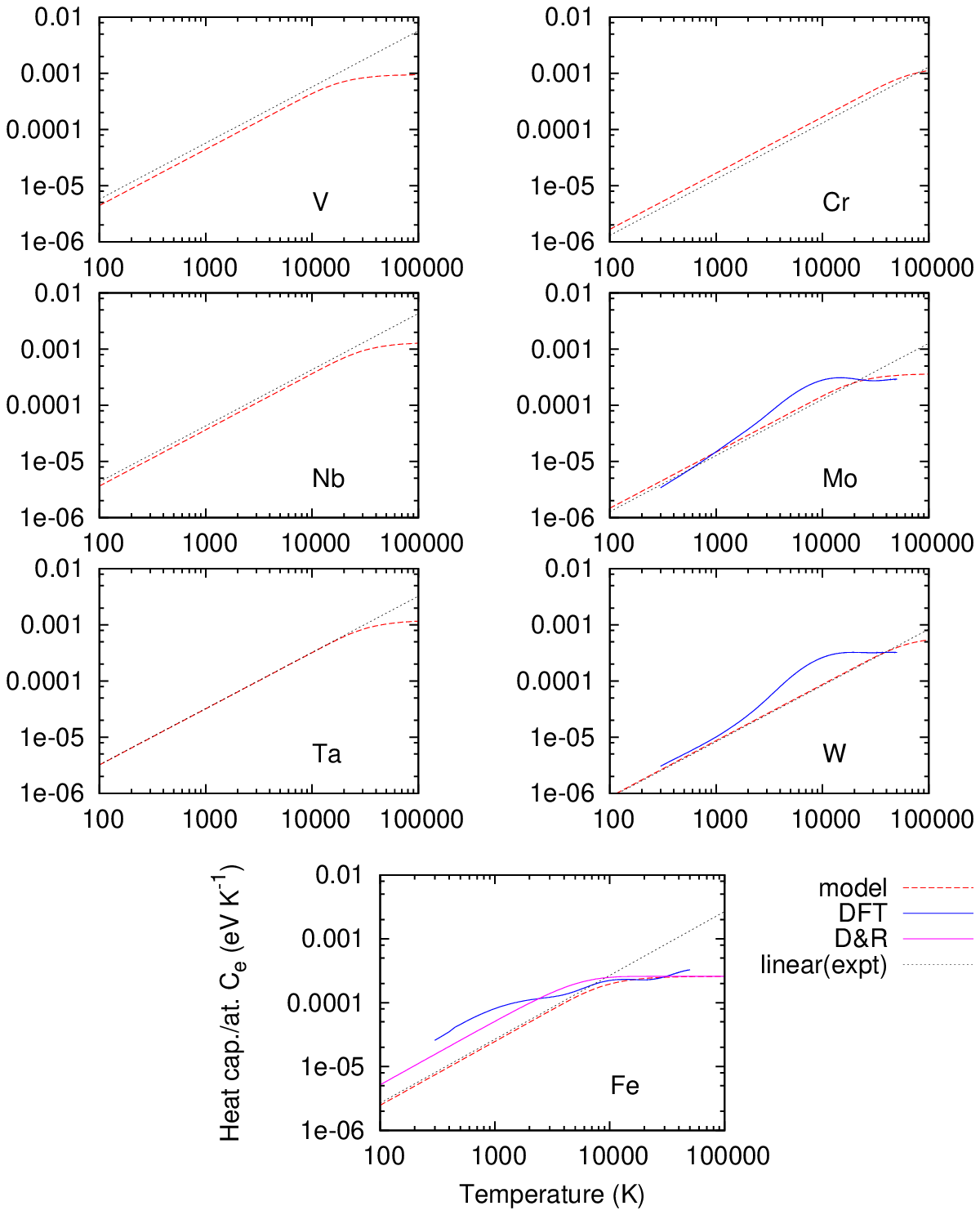}
           \end{center}
           \caption {
           Electronic heat capacity per atom  as a function of temperature.
           The dotted line is the linear function $C_e = \gamma T_e$, an extrapolation from the low-temperature linear experimental heat capacity from ref\cite{Tari_2003}.
           Note that at cryogenic temperatures heat capacity is affected by electron-phonon coupling, so $\gamma = 1/3 \pi^2 k_B^2 T_e D ( 1 + \lambda )$ is actually measured \cite{Thiessen_IJTh1985}.
           The factor $( 1 + \lambda )$ has therefore been removed to make a comparison above the Debye temperature.
           The solid blue line are the DFT estimates of Lin\cite{Lin_PRB2008}. Note they use spin-degenerate iron. 
           Lines for EAM AT and DND parameterizations are almost indistinguishable.
           The line marked D\&R is the heat capacity used by Duffy \& Rutherford\cite{Duffy_jpcm2007}.
            }
           \label{heatcapacity}
       \end {figure}

        The electronic heat capacity per atom must be the derivative of $\Theta$ with respect to temperature, which at low temperature is 
            \begin{equation}
                \lim_{T_e/W_i\rightarrow 0} C_{e,i} =  \frac{\pi^2 k_B^2 T_e}{3}  \left(\frac{2 N_a}{W_i}\right).
            \end{equation}
        We can fit the density of states at the Fermi level to that found by a DFT calculation. 
        Then the heat capacity of the perfect lattice is material-dependent but potential-independent; its variation with environment is potential-dependent.
        Electronic heat capacity is plotted in figure \ref{heatcapacity}.
        
	\section{Thermal conductivity}
	
		Energy stored due to a raised electronic pseudotemperature $\Theta_{ T_{e},i }$ can diffuse away by electrical thermal conductivity.
		The kinetic theory expression for the thermal conductivity is written in terms of the Fermi electron velocity $v_F$, scattering time $\tau_e$ and electronic heat capacity per unit volume $C_{e}=1/\Omega \, \partial \Theta / \partial T_e$ as 
			\begin{equation}
				\kappa = \frac{1}{3} \langle v_F^2 \rangle C_{e} \tau_e.
			\end{equation}
		
        The Fermi velocity can be computed by DFT. Our calculations are summarised in \ref{FermiVelocityCalc}.

		\subsection{Electron scattering rate}
		
		The electron scattering rate is given by the sum of impurity, electron-phonon and electron-electron scattering rates according to Mattheisen's rule\cite{MottJones_1936}
			\begin{equation}
				\frac{1}{\tau_e} =    \frac{1}{\tau_{e-e}} + \frac{1}{\tau_{e-ph}} + \frac{1}{\tau_{\mathrm{imp}}}
			\end{equation}
		The electron-electron scattering rate will be proportional to the number of thermal electrons and holes, so $\frac{1}{\tau_{e-e}} \sim T^2$.
		At low temperature the electron-phonon scattering rate will be proportional to the number of phonons, so $\frac{1}{\tau_{e-ph}} \sim T^3$,
		and in the high temperature limit (above the Debye temperature) the number of phonons capable of scattering electrons $\sim T$, so $\frac{1}{\tau_{e-ph}} \sim T$.
		Mott-Jones impurity scattering will be dependent on the degree of disorder/ concentration of defects in the system, and be largely temperature independent.
		
		Electron-electron scattering is a quasiparticle effect beyond the scope of this paper. 
		We will fit $\frac{1}{\tau_{e-e}} = \sigma_2 T^2$ to experimental thermal conductivity data.
		
		\subsection{Electron-phonon scattering}
		\label{eph_scattering}								
		We can find an environmentally dependent form for the electron-phonon scattering as follows.
		
		At high temperatures and thermal equilibrium Kaganov proposed the following for heat transfer per unit volume from ions to electrons\cite{Kaganov_JETP1957}
			\begin{equation}
				\left. \frac{\partial E_e}{\partial t} \right|_{e-ph} = G_K \, T_I, \quad G_K = \frac{\pi^2}{6} \frac{ m_e c_s^2 N_e }{ \Omega \tau_{e-ph} T_e },
			\end{equation}
		where $c_s$ the speed of sound.
		Equating $G_K = G_{\mathrm{EAM}}$ from equation \ref{GEAM} at $T_I=T_e$ gives an environmentally dependent electron-phonon scattering rate
			\begin{equation}
				\label{electronphononrate}
				\frac{1}{\tau_{e-ph}} = \frac{  \bar{W} }{W_i} \sigma_1 \bar{B}_i T_I,
			\end{equation}
		where $\sigma_1 \equiv \frac{ 18 k_B }{ M_I m_e c_s^2 N_e }$ is a material constant given in table \ref{parameterization} using mass and speed of sound from ref \cite{webelements}.
		The requirement that the scattering time from equation \ref{electronphononrate} reproduces the temperature variation of experimental thermal conductivity defines a constraint on the model electrons per atom $N_e$. 
		
		It should be noted that one element, iron, does not give a sensible value for the number of electrons per atom using this constraint- indeed it would require $N_e=81.96,N_a=41.04$ to do this.
		This is an indication that iron may not be reasonably parameterized to reproduce both experimental thermal conductivity and experimental stopping power by this model.
		The key physics absent in this simple model is spin-dependence, and an empirical fit for $\sigma_1$ in iron is instead given in table \ref{parameterization}.
		A better model for iron which includes the full spin-fluctuation dissipation mechanism has been given by Ma \& Dudarev \cite{Ma_PRB2011,Ma_PRB2012}.
		Our inclusion of an iron parameterization within such a simplified model is intended to allow researchers to make quick comparisons between iron and other bcc metals.
		
		\subsection{Electron-impurity scattering}
		\label{eimp_scattering}					
		The Mott-Jones impurity scattering time in the dilute impurity limit must take the form\cite{MottJones_1936}
			\begin{equation}
				\frac{1}{\tau_{\mathrm{imp}}} = \frac{v_F}{\lambda_{\mathrm{Mott}}} = \frac{ N_{\mathrm{impurity}} }{V} v_F \langle \sigma(k_F) \rangle,
			\end{equation}
		where $N_{\mathrm{impurity}}/V$ is the number of impurities per unit volume and 
			\begin{equation}	
				\langle \sigma(k_F) \rangle = \int \left[ 1 - \cos \theta \right] \sigma( k_F,\theta ) d\Omega_s,
			\end{equation}
		is the average scattering cross section at the Fermi wavevector.
		For a full calculation differential scattering cross sections may be determined from the transition matrix elements coupling electronic states.

		For the purposes of MD a calculation of differential scattering cross sections is impractical.
		However we can make an estimate for the scattering rate for lattice defects in a pure metal as follows:
		The electron-phonon scattering rate given in equation \ref{electronphononrate} gives the scattering by the motion of ions linear in temperature.
		Impurity scattering must make an additional contribution for athermal atoms, and to detect these within an MD simulation suggests using a function of energy,
			\begin{equation}
				\frac{1}{\tau_{\mathrm{imp},i}} = \frac{v_F}{a_0} s( E_i ),
			\end{equation}
		where $s(E_i)$ is non-zero for athermal atoms.
		
		The simplest possible model is to write that the scattering rate is directly proportional to the degree of disorder, as measured by the potential energy.
			\begin{equation}
				\frac{v_F}{a_0} s( E_i ) = \left\{ \begin{array}{lr}
						\sigma_0 \frac{W_i}{\bar{W}} \Delta E_i	&	\Delta E_i > 3 k_B T_e			\\
						0																&	\mbox{otherwise},
						\end{array} \right.
			\end{equation}
		where $\Delta E_i = V_i + F_{T_e,i} - E_{\mathrm{coh}}$ is the surplus potential energy.
		
		We can use the Wiedemann-Franz law ($r = L T/\kappa $, where $L = 2.44\times 10^{-8}$ W$\Omega$K$^{-2}$ is the Lorenz number )to relate electrical resistivity $r$ to given thermal conductivity.
		Hence we can fit the constant $\sigma_0$ to experimental values for the electrical resisitivity per Frenkel pair $r_{\mathrm{FP}}$, giving the table \ref{MottScattering}, 
			\begin{equation}
			    \label{impurityScattering}
				\sigma_0 = \frac{ r_{\mathrm{FP}}} {H^F_{\mathrm{FP}}} \frac{ v_F^2 C_e} {3 L \Omega T_e}.
			\end{equation}
		It should be noted that where experimental resistivities per vacancy are available, this very simple model gives good values.
		$\sigma_0$ is material dependent, but not potential dependent.

		\begin{table}
	        \centering
	        \small
	        \begin{tabular}{l|ll|ll|ll|ll}	  
	                    &                           &               &   \multicolumn{2}{c|}{expt}                 			&   \multicolumn{2}{c|}{AT}                   &   \multicolumn{2}{c}{DND}   \\
	        	Element	&	$H^f_{\mathrm{FP}}$	    &	$\sigma_0$	&   $r_{\mathrm{FP}}$	&	$r_{\mathrm{V}}$	&		$r_{\mathrm{FP}}$	&	$r_{\mathrm{V}}$	  &	$r_{\mathrm{FP}}$	&	$r_{\mathrm{V}}$	          \\
	        			&	(eV)				    &		    	&   $\mu \Omega$ m/at.fr.   &   &&  &&             	      	\\
	        	\hline                         	           			                                                                                                                                                                           	
	        	V       &	  5.877            	   	&   4.92		&   21                  	&	                	    &		15.8             		&	6.0       &	 13.3			&	7.9     	                \\
	        	Nb      &	  8.243            	   	&   3.14   		&   14                  	&	                	    &		10.8            		&	4.1       &	 9.6			&	4.4     	                \\
	        	Ta      &	  8.976            	   	&   2.98   		&   16.5               		&	               		    &		14.7            		&	5.4       &	 10.9		    &	5.1    		                 \\
	        	Cr      &	  8.3              	   	&   4.45   		&   37                  	&	                	    &		17.9         			&	7.3       &	  				&	        	                \\
	        	Mo      &	  10.139           	   	&   2.07   		&   13.4                	&	4.3                	    &		9.8             		&	3.1       &	 11.0 			&	3.9        	             \\
	        	W       &	  13.111           	   	&   2.22   		&   27                  	&	7                	    &		20.6            		&	6.8       &	 19.3		    &	6.7      	                \\
	        	Fe		&	  6.08             	   	&   4.55        &   24.6                	&	                	    &		6.8             		&	8.4       &	              	&	            	                
	        \end{tabular}                                                                                                                                                                                                                               
	        \caption{                
	        	A simple model for Mott-Jones impurity scattering: scattering cross section proportional to defect formation energy parameterised for the pure metals.
	        	Frenkel pair formation energies from ref\cite{Derlet_PRB2007}.
	        	Experimental resistivities per Frenkel pair from recommended values in ref\cite{Broeders_JNM2004}, resistivity per vacancy from ref\cite{Landholt-Bornstein}.
            	Model resistivities are computed using the standard Ackland-Thetford / Derlet-Nguyen-Manh potentials at 1K in a cell of 1008$\pm$1 atoms at constant volume.
            	The interstitial configuration used is the $<111>$ crowdion for all elements except iron ($<110>$ dumbbell).
	        }
	        \label{MottScattering}
	    \end{table}

			To complete this section we must consider the Ioffe-Regel limit where the electron scattering is high, and so thermal conductivity is low\cite{IoffeRegel_ProgSemi1960}.
			Physically the electron mean-free path can not be much smaller than the separation between atoms.
			For a bcc metal this separation is $\sqrt{3/4}a_0$, so this suggests $1/\tau_e < v_F/(\sqrt{3/4}a_0)$.
			It has been shown experimentally and theoretically that a reasonable model is that electrical resistivities add in parallel\cite{Wiesmann_prl1977}
			$1/r = 1/r_{max} + 1/r(T)$, so we suggest the final form for electron scattering time
				\begin{eqnarray}
					\label{scatteringTime}
					\tau_{e,i} &=& \frac{\sqrt{3/4}a_0}{v_F} + \left( \frac{1}{\tau_{\mathrm{imp}}} + \frac{1}{\tau_{e-ph}} + \frac{1}{\tau_{e-e}} \right)^{-1} 	\nonumber 	\\
						&=&	\frac{\sqrt{3/4}a_0}{v_F} + \left\{ \begin{array}{lr}
								\left( \sigma_0  \frac{W_i}{\bar{W}} \Delta E_i 
								+ \sigma_1  \frac{  \bar{W} }{W_i} \bar{B}_i T_I 								
								+ \sigma_2 T_e^2							
								\right)^{-1}		&	\Delta E_i > 3 k_B T_e		\\
								\left( \sigma_1  \frac{  \bar{W}}{W_i} \bar{B}_i T_I 
								+ \sigma_2 T_e^2							
								\right)^{-1}  		&	\mbox{otherwise}
								\end{array} \right.
				\end{eqnarray}

\begin{figure}[h!t!b]
           \begin {center}
               \includegraphics [width=5.0in] {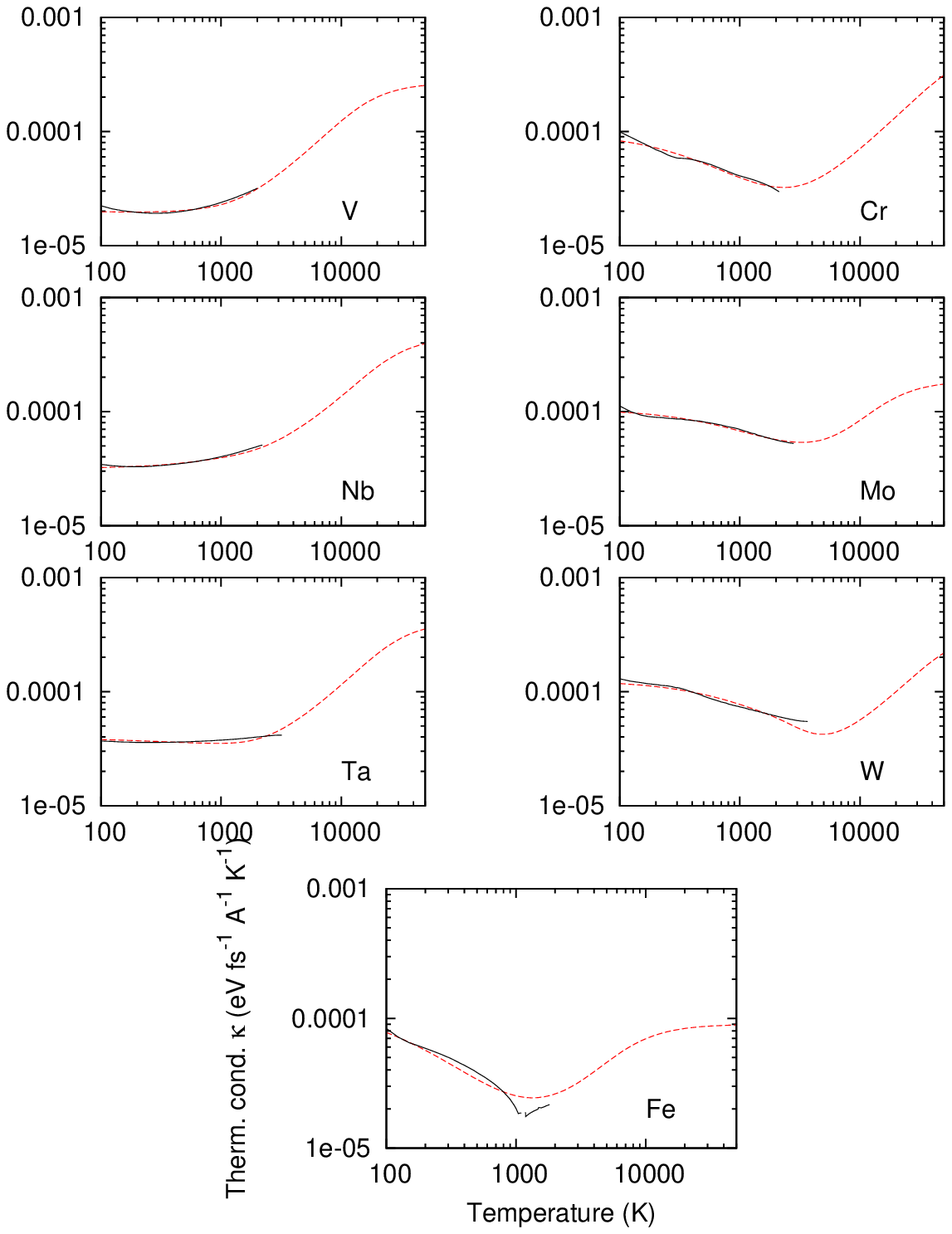}
           \end{center}
           \caption {
           Electronic thermal conductivity per atom  as a function of temperature.
           The solid line is experimental data from ref\cite{Ho_JPCRD1972}.
           Lines for EAM AT and DND parameterizations are almost indistinguishable.
            }
           \label{thermalConductivity}
       \end {figure}

		We can fit the temperature dependence of the scattering rate to the experimental thermal conductivity of the pure metal.
		This is done using linear-least-squares fitting assuming all the measured conductivity is due to the electrons, there is no thermal expansion, and electrons and ions are in equilibrium.
		The fit is performed in the range 100K to the metal's melting point.
		Athermal defect scattering is assumed using the model in equation \ref{impurityScattering}, with the expected concentration of atoms with energy $\Delta E$ proportional to $\exp[-\Delta E/k_BT]$.
		No other lattice defects are included.
		The results are shown in figure \ref{thermalConductivity}.

	\section{Two-temperature Langevin dynamics}
		\label{SPE}

		To ensure that an equilibrium is found we must use the fluctuation-dissipation theorem for dissipative particle dynamics\cite{Espanol_epl1995}, in this case we use a random Gaussian white noise return force acting on bond $i-j$, along the direction $\mb{e}_{ij} = \frac{\mb{R}_{ij}}{|\mb{R}_{ij}|}$.
		This random force we write as $\xi_{ij}$
			\begin{eqnarray}
			    \label{twotmd_eph_force}
				\mb{F}_{\mathrm{e-ph+SPE},i} &=& \frac{  \bar{W} }{W_i} \sum_{j \in \mathcal{N}_i} \, \mb{B}_{ij} ( \dot{\mb{R}}_i-\dot{\mb{R}}_j )
											+ \sum_{j \in \mathcal{N}_i} X_{ij} \mb{e}_{ij}  \xi_{ij},					\nonumber	\\
				\mbox{where} \quad
											X_{ij}^2 &=& 2 k_B T_{e,i} \frac{  \bar{W} }{W_i} \mb{e}_{ij}^T  \mb{B}_{ij} \mb{e}_{ij}
			\end{eqnarray}											
		
		We can now complete our prescription for two-temperature molecular dynamics.
		The total force on atom $i$ is given by 
			\begin{equation}
			    \label{twotmd_force}
				\mb{F}_i = -\nabla_i V - \nabla_i F_{0,i} - \nabla_i \Theta_{ T_{e},i } + \mb{F}_{\mathrm{e-ph},i} + \mb{F}_{\mathrm{SPE},i}.
			\end{equation}
			
		The thermal conductivity in the neighbourhood of atom $i$ is given by 
			\begin{equation}
				\kappa_i = \frac{v_{F,i}^2}{3} C_{e,i} \tau_{e,i},
			\end{equation}
		where $\tau_{e,i}$ is given by equation \ref{scatteringTime}, and $v_{F,i}$ is given by equation \ref{localFermiVelocity}.
		The thermal conductivity for a volume $\Omega_S$ containing a set of atoms $ \{1,2,3,\ldots,N_S\}$ is given by
			\begin{equation}
				\frac{1}{\kappa_S} = N_S \sum_{i \in \Omega_S} \frac{1}{\kappa_i},
			\end{equation}	
		and the electronic temperature in $\Omega_S$ evolves as 			
			\begin{equation}
			    \label{twotmd_conduction}
				\left( \sum_{i \in \Omega_S} C_{e,i} \right)  \frac{ \partial T_{e} }{\partial t} = \nabla \cdot ( \kappa_S \nabla T_{e} )   - \sum_{i \in \Omega_S} \left( \mb{F}_{\mathrm{e-ph},i} \cdot \dot{\mb{R}}_i + \Omega G_{\mathrm{EAM}} T_e \right),
			\end{equation}
        where we have avoided stochastic calculus by replacing the explicit energy transfer from the stochastic return with its ensemble average value. 
        
        We find that to compute damping, heat capacity, and thermal conductivity on the fly for this full formalism takes order 3 times longer calculation time than straightforward energy conserving MD: on a single core on a desktop PC 11s per million atom update steps for the full model compared to 3.5s per million atom update steps.
        Most of this additional time is taken by replacing the Velocity Verlet integrator with an adaptive timestepping 4th order Runge-Kutta algorithm.
        The diffusion term can be evaluated stably and efficiently using a grid of electron cells with a 27-point stencil\cite{OReilly_JINM2006}. 
        This allows us to take a timestep of up to 2fs after the point of peak damage.

	\begin{table}
        \centering
        \small
        \begin{tabular}{l|l|lllllll}
        	    &   Element		                    &	  V		&	Nb		&	Ta		&	Cr		&	Mo		&	W		&	Fe               \\      
        	    \hline                                                                                                     
			    &   $N_e$		                    &	11.126	&	15.380	&	14.080	&	15.431	&	4.243	&	7.279	&	3.000            \\            
			    &   $\sigma_0$(fs/eV)               &   4.92 	&	3.14	&	2.98  	&	4.45  	&	2.07  	&	2.22   	&	4.55        	\\                        
			    &   $\sigma_1$                      &   225.0   &	152.2   &	90.1	&	103.3  	&	162.2  	&	77.3   &	8.01$^*$        \\              
			    &   $\sigma_2$(10$^{-9}$ fs/K$^2$)	&	844.6	&   124.3 	&	364.7	&	309.8	&	170.6	&	95.9	&	892.1		\\                 
			    &   $v_F$ (\AA/fs)					&	4.47	&	6.46	&	6.47	&	5.73	&	8.72	&	9.50	&	4.95			        \\                 
		\multirow{2}{*}{FS} 	    &   $N_{a,FS}$		                &	6.674	&	8.667	&	7.985	&	7.908	&	3.346	&	4.155	&	3.032            \\            
			    &   $\zeta_{FS}$		            &   0.0863 	&	0.0857	&	0.1089	&	0.2114	&	0.0463	&	0.0385	&	1.9895           \\            
		\multirow{2}{*}{DND}	    &   $N_{a,DND}$		                &	6.448 	&	8.590 	&	7.971	&	 	    &	3.212 	&	3.294 	&	                \\             
			    &   $\zeta_{DND}$		            &   0.8224	&	0.8604	&	1.7159	&		    &	0.9468	&	1.2146	&	                \\             
			    \hline                                                                                                                                     
			    &   $a_0$ (\AA) 					&	3.0399	&	3.3008	&	3.3058	&	2.8845	&	3.1472	&	3.1652	&	2.8665                   \\                      
			    &   $E_{\mathrm{coh}}$ (eV) 			&	5.31	&	7.57	&	8.10	&	4.10	&	6.82 	&	8.90	&	4.28                   \\                      
		\multirow{3}{*}{FS} 	    &   $w_{FS}$	(eV)	            &	2.170  	&	3.476	&	3.110	&	7.738	&	2.430	&	4.198 	&	2.413			\\                           
			    &   $\bar{W}_{FS}$(eV)	            &	7.375	&	11.634 	&	12.190	&	23.156	&	11.153	&	23.438	&	6.003			\\
			    &   $W'_{FS}$		                &   4.787   &	5.690   &	5.566   &   11.769	&	3.622  	&	5.545   &   3.535 		\\                       
		\multirow{3}{*}{DND}	    &   $w_{DND}$	(eV)	            &	7.125	&	11.531	&	12.170	&		    &	10.705	&	18.579	&				\\
			    &   $\bar{W}_{DND}$(eV)	            &	7.125	&	11.531	&	12.170	&		    &	10.705	&	18.579	&				\\
			    &   $W'_{DND}$		                &   5.208   &	5.883   &	5.599   &		    &	3.687  	&	5.189   &				\\                       
        \end{tabular}                                                                                                                                                                                                                                                                                                      
        \caption{                
        	Complete parameterization for the elements considered.
        	The units of $\sigma_1$ are $10^{-6}$ A$^2$/eV/fs$^2/K$.
        	$^*$Iron may not be fit to both experimental thermal conductivity and electron-phonon stopping using the formulae of this simple model.
        	For $N_e=3$, $\sigma_1$ should be 65.1: the value tabulated here is our recommended empirical fit (see text).        	
        	The first set of parameters ($N_e$ to $\zeta$) define the non-adiabatic correction.
        	The second set of parameters are derived quantitiese.
	        }
        \label{parameterization}
    \end{table}
%
       
    \subsection{2TMD simulations}
        To test the importance of the heat transfer model used, we have performed some small-scale illustrative simulations of cascades using 2TMD.
        We have simulated the displacement cascade formation and recombination of 25keV PKA W self-ions in W, using the temperature dependent Ackland-Thetford potential described here.
        \footnote{Following Fikar \& Sch\"aublin\cite{Fikar_JNM2009}, we tweak the original potential by switching the repulsive part of the potential to the universal form proposed by Ziegler et al.\cite{Ziegler:1985fj} between 1.0A and 1.5A, using a simple fifth-order polynomial.}
        It is not our intention in this paper to draw out the difference in metastable defect populations due to the heat-transfer model used, as while this is perhaps the most technologically significant quantity, it requires a large number of independent runs to gather suitable statistics, and then would necessitate disambiguating effects of the potential form from those of the heat transfer.
        Instead our focus is on the dynamic partition of the energy between ionic kinetic, potential, and electronic thermal energy.
        
        For each heat transfer model we have used the same set of four statistically independent starting configurations, starting with a system of 64$^3$ unit cells (524288 atoms) well-thermalised to 300K with periodic boundary conditions. The cascades were started near the centre of the box so to not overlap the periodic boundary.
        32$^3$ electronic temperature cells were used with periodic boundary conditions, with each electronic cell containing $16^3$ atomic unit cells (8192 atoms).
        
        The models used are:
        \begin{description}
            \item{NVE}\newline
                The simulation was performed in a thermally isolated supercell so that internal energy (KE+PE) is conserved.
            \item{NVT}\newline
                A Langevin thermostat was added (ie using equation \ref{twotmd_force}), with the damping force given by $\mb{F}_{\mathrm{e-ph},i}=-\bar{B}_0 \dot{\mb{R}}_i$, with $\bar{B}_0=1.188$ the damping constant for an atom in the perfect lattice. The electronic temperature is fixed, equivalent to taking the limit of infinite electronic thermal conductivity and heat capacity.
            \item{Kinetic energy cutoff $k_c=1.0eV$}\newline
                Atoms with a kinetic energy $\half M_i | \dot{\mb{R}}_i |^2 > 1.0$eV were damped with a force $\mb{F}_{\mathrm{e-ph},i}=-\bar{B}_0 \dot{\mb{R}}_i$, with no stochastic returning force, ie $\mb{F}_{\mathrm{SPE},i}=0$. 
                To produce a meaningful temperature in the long-time-limit, all atoms within two unit cells of the boundary were thermalized using the Langevin thermostat described above.
                This model is intended to be similar to that used by Sand et al.\cite{Sand_EPL2013}.
            \item{Kinetic energy cutoff $k_c=10.0eV$}\newline
                As above, but only those atoms in the central region with kinetic energy $>10.0$eV damped.
            \item{2TMD with fixed damping and thermal conductivity a function of temperature only, ( 2TMD$(\bar{B},\bar{\kappa})$ )}\newline
                The damping force is given by $\mb{F}_{\mathrm{e-ph},i}=-\bar{B}_0 \dot{\mb{R}}_i$, with $\bar{B}_0=1.188$ and a stochastic Langevin return.                
                Heat transfered to the electrons then diffuses according to the heat-diffusion equation, with fixed thermal conductivity $\bar{\kappa}(T_e)$ correct for the perfect lattice at the local electronic temperature.
                This model is intended to be similar to that used by Duffy and Rutherford\cite{Duffy_jpcm2007}.
            \item{2TMD, using the full formalism described here}\newline
                The damping and stochastic return force is given by equation \ref{twotmd_eph_force}, and the thermal conductivity computed locally within the electron cells.                                          
        \end{description}
        
\begin{figure}[h!t!b]
          \begin {center}
              \includegraphics [width=4.0in] {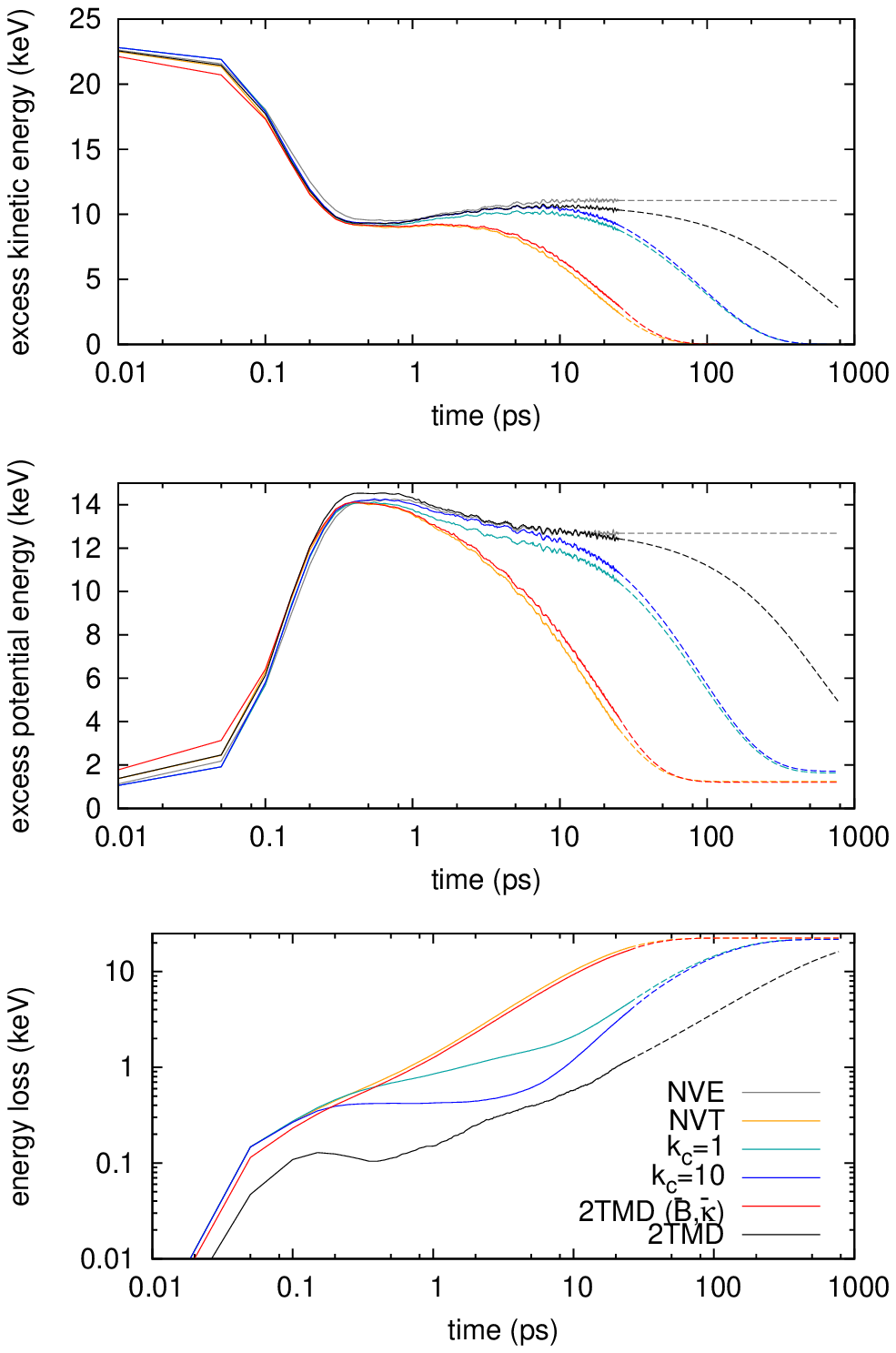}
          \end{center}
          \caption {
          Results of MD simulations of 25keV cascades in tungsten using different models for heat transfer.
          Results are averaged over four independent runs. The simulations were terminated after 25ps, and the heat transfer after this point extrapolated by solving equation \ref{heatTransferEqn} (dotted lines).
          Energy loss is defined as the difference in ionic energy (kinetic+potential) within the MD simulation box at time $t$ and time $0$. Note that NVE,2TMD$(\bar{B},\bar{\kappa})$ and 2TMD simulations are energy conserving to within a few volts over this timescale if the energy of the electron cells is taken into account.
           }
          \label{tungsten2TMDExample}
        \end {figure}

        Results are shown in figure \ref{tungsten2TMDExample}.        
        We see the excess kinetic energy ( the 25keV given to the PKA ) is rapidly converted to potential energy at the position of maximum damage, about 0.8ps into the simulation.
        The NVE simulation subsequently partitions this excess energy between ionic potential and kinetic energy, raising the effective ionic temperature within the box to 460K.
        The other simulations have some form of thermostat, and so return ionic temperature to 300K.
        
        The rate at which this occurs is dependent on the model used.
        It can be seen that the NVT and 2TMD$(\bar{B},\bar{\kappa})$ models give very similar results in this test.
        This is because crystalline tungsten has a very high electronic thermal conductivity, so any electronic temperature spike is rapidly diffused away.
        
        The kinetic energy cutoff models show an initial fast transfer of energy from the MD region to the electrons, but this then plateaus when the ions are slowed to below the kinetic energy cutoff. 
        After 10ps lattice vibrations in the ions have reached the thermalizing boundary and so energy is lost again.

        It was necessary to terminate the simulations at 25ps, owing to the small size of the simulation box.
        However it is possible to extrapolate the further heat transfer from ions to electrons using equation \ref{heatTransferEqn}, assuming that the ionic kinetic and potential energy is equipartitioned at this point.
        For the NVT and 2TMD$(\bar{B},\bar{\kappa})$ models, this simply means using the perfect lattice damping constant $\bar{B}_0$.
        For the kinetic energy cutoff models, $\bar{B}_0$ is again used, but now heat is only exchanged by boundary atoms.
        It can be seen in figure \ref{tungsten2TMDExample} that the heat flow gradients are accurately matched by this simple model.
        
        For the full 2TMD model it is harder to extrapolate exactly beyond 25ps.
        The quotient of ionic energy loss rate and ionic kinetic energy was averaged over the last 5ps of simulation, and this value used to extrapolate kinetic energy. As we should expect the quotient itself to evolve with the ionic and electronic temperature, the extrapolation in figure \ref{tungsten2TMDExample} should be treated as an indication of the likely relative position of the model curves only.            
        
\begin{figure}[h!t!b]
          \begin {center}
              \includegraphics [width=4.0in] {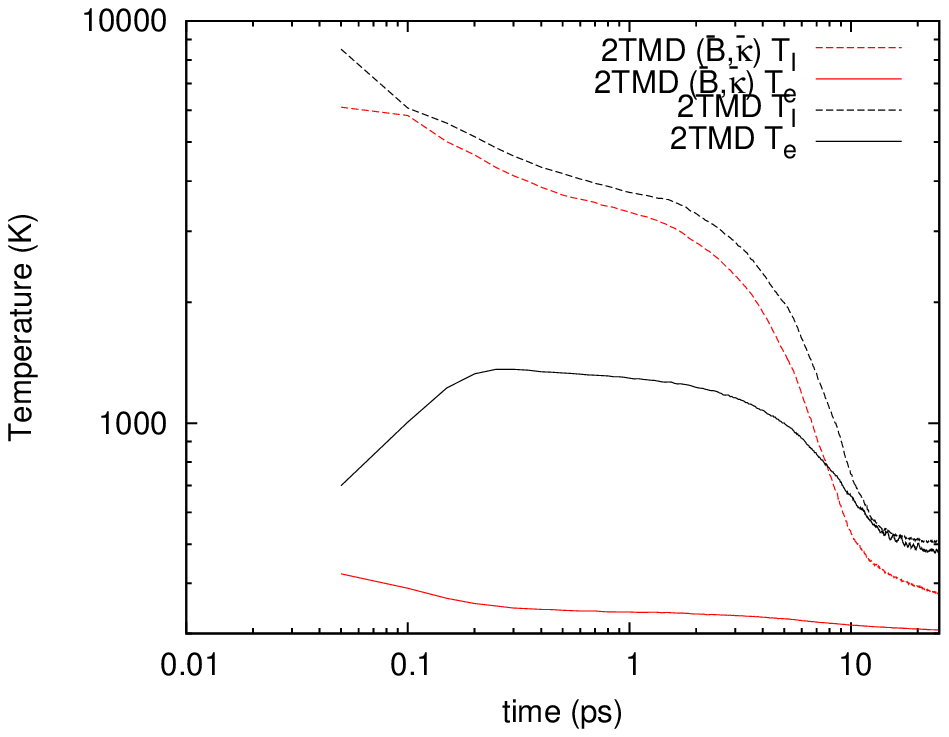}
          \end{center}
          \caption {
          Maximum ionic and electronic temperature recorded in an electron cell in 2TMD simulations of 25keV cascades in tungsten, with and without the assumption of thermal conductivity a function of temperature only.
          Note that at short times the ions are far from being in thermal equilibrium, so the y-axis should more properly be interpretted as a pseudotemperature.
          Each electron cell contains an average of 8192 atoms.
           }
          \label{tungstenTemperature}
        \end {figure}

        The full 2TMD model shows strikingly different behaviour to the other heat-transfer models.        
        As the displacement cascade evolves, the electronic thermal conductivity is greatly reduced, and so the electronic cells containing the cascade stay much hotter. 
        This can be seen in figure \ref{tungstenTemperature}.
        The ions and electrons are closer in temperature, so less net heat transfer is seen at the point of maximum damage.
        After 10ps evolution, as the defected ions recombine and cool, the ions and electrons are close to being in thermal equilibrium.
        
        It is notable that the energy loss for 2TMD and $k_c=10$eV models are comparable at 10ps.
        On the one hand the 2TMD model used here raises electronic temperature and so reduces the energy loss from all ions, on the other hand the $k_c=10$eV only takes energy from the ions in the early part of the cascade.
        It is possible that the undoubted success of using a kinetic energy cutoff\cite{Sand_JNM2015} is related to this observation that it gives a good estimate of the energy extracted from the cascade.

	\section{Discussion and Conclusion}
	
		We have derived simple analytical formulae for the non-adiabatic properties of bcc metals suitable for use in Molecular Dynamics simulations, 
		and parameterized them for two well-known sets of empirical potentials, namely the Ackland-Thetford and Derlet-Nguyen-Manh-Dudarev potentials.
		It is notable how similar the corrections are for these sets.
		
		The number of electronic d-band states ($N_a$) assumed within the models are slightly different.
		$N_a$ is fitted to reproduce the bond energy, which while it may show differences from one EAM potential to the next, is constrained to lie within a fairly tight range.
		The number of electrons ($N_e$) is exactly the same, which is required to reproduce the same electron-phonon scattering rate.
		
		We have introduced an order-1 fitting parameter ($\zeta$) to tune the electron-phonon stopping factor to the experimentally determined value.
		This shows differences between potentials, as it balances the squared gradient of the electron density with respect to atomic separation.
		This gradient is not generally a fitted parameter for an EAM potential: it is usually only considered in combination with the gradient of the repulsive part of the potential.
		
		The heat capacity of the metal is defined by the change in bond energy as a function of temperature.
		As we fit the heat capacity of the metal at low temperature to the DFT value, the scattering time constants are necessarily equal across potentials in order that they reproduce the same temperature variation.
		
		We can therefore suggest with some confidence that the corrections contained in these parameters are a reasonable guess for other EAM potentials not covered in this paper.
		Note that it will be necessary to choose either the FS or DND forms for the electron density to generate the electronic stopping tensor unless $\zeta$ is refitted appropriately.

		We have shown how to compute the electron-phonon and electron-electron scattering times for use within MD simulation, and have also proposed a very simple form for impurity scattering in a pure but defected metal.		
		While we should expect the impurity scattering formula to be improved subsequently; our simple forms have the correct physical content and correct magnitude and can be therefore be replaced as knowledge improves.
		Importantly we offer a form for impurity scattering which should give a sensible result for the sparsely defected limit where only a few Frenkel pairs are present, a sensible result for the densely defected limit where the Ioffe-Regel limit is invoked, and can identify the boundary beween these regions on-the-fly.

        We have demonstrated the use of these formulae with some example 2TMD simulations of 25keV collision cascades in tungsten.
        We see that the reduction in electronic thermal conductivity due to disorder within the lattice leads to a dramatic increase in electronic temperature during cascade formation, and the heat is retained within the electrons for a much longer time.
        This will surely lead to significant differences in predictions of residual damage in irradiated metals compared to earlier fixed conductivity models.

	\section{Acknowledgements}

		This work has been carried out within the framework of the EUROfusion Consortium and has received funding from the Euratom research and training programme 2014-2018 under grant agreement No 633053. The views and opinions expressed herein do not necessarily reflect those of the European Commission. To obtain further information on the data and models underlying this paper please contact PublicationsManager@ccfe.ac.uk.

\appendix

\section{Electronic stopping in the tight-binding approximation with explicit electrons}
    \label{StoppingCalc}			
		We write the positions of $N$ ions at time $t$ as $\mb{R} = \{ \mb{R}_1(t),\mb{R}_2(t),\ldots \mb{R}_N(t)\}$, and the electrons are represented by a single particle density matrix $\hat{\rho}(t)$ which evolves according to a time-dependent potential due to the position of the ions.
		If we neglect electron-electron interactions the energy for this system is given by the Harris-Foulkes Tight-Binding functional form\cite{HarrisFoulkes}:
			\begin{equation}
				E\left( \mb{R} ; t \right)_{TB} = \half \sum_i M_i \left| \dot{\mb{R}}_i \right|^2 + V\left( \mb{R} \right) + 2\, \mathrm{Tr} \left[ \hat{H}\left(\mb{R}\right) \, \hat{\rho}(t) \right],
			\end{equation}
		where $M_i$	is the mass of the $i$th ion, $V\left( \mb{R} \right) = \half \sum_i \sum_{j \in \mathcal{N}_i} V\left( \mb{R}_i - \mb{R}_j \right)$ represents the classical ion-ion interaction energy as a sum over pairs of ions, and the 2 is for spin degeneracy.
		
		Dynamic evolution of electrons and ions which conserves energy is then given by 	
			\begin{eqnarray}
				M_i \ddot{\mb{R}}_i &=& -\nabla_i V\left( \mb{R} \right) - 2\, \mathrm{Tr} \left[ \nabla_i\hat{H}\left(\mb{R}\right) \, \hat{\rho}(t) \right],	\nonumber \\
				i \hbar \dot{\hat{\rho}}(t) &=& \left[ \hat{H}\left(\mb{R}\right),\hat{\rho}(t) \right].
			\end{eqnarray}
		The ions therefore move according to their Hellman-Feynman forces, and the electron evolve according to both the instantaneous position of the ions and the history of their motion.
		Note that if the electrons start in a ground state, the only possibility is that they gain energy, and the ions lose energy.

		If the eigenvalues of the instantaneous electron-ion Hamiltonian at time $t$ are 
			\begin{equation}
				\hat{H}\left(\mb{R}\right) |m\rangle = \epsilon_m |m\rangle, 
			\end{equation} 			
		then for a well thermally-equilibrated system the electron density matrix will be close to the canonical Born-Oppenheimer form
			\begin{equation}	
				\hat{\rho}_{T_e}\left(\mb{R}\right) = \sum_m f\left( \epsilon_m ; \lambda,T_e \right) |m\rangle \langle m|,
			\end{equation}
		where $f\left( \epsilon ; \lambda,T_e \right) = \left(1+ \exp{ \left[(\epsilon-\lambda)/k_B T_e\right] } \right)^{-1}$ is the Fermi-Dirac function for electronic temperature $T_e$ and Fermi energy $\lambda$.
		Note that $\hat{\rho}_{T_e}\left(\mb{R}\right)$ is explicitly dependent on the atomic positions, whereas $\hat{\rho}(t)$ is not.
	
		If ions and electrons are out of equilibrium it is still possible to describe the density matrix using a pseudotemperature, and its deviation from this canonical Born-Oppenheimer form.
		Indeed in radiation damage cascades some 95\% of the change in the forces on the ions due to electronic stopping are captured by such a simple pseudotemperature model\cite{Race2009}.
		The time-evolved density matrix can therefore be written as the density matrix appropriate for a canonical pseudotemperature plus a correction.
		As the effect of the long-term history of ion movements is to raise the pseudotemperature it follows that if the ions are moving slowly compared to the electrons, the principal part of this correction must be time-local: 
			\begin{equation}	
				\rho(t) = \hat{\rho}_{T_e}\left(\mb{R}\right) - \tau\, \dot{\mb{R}} \cdot \nabla \hat{\rho}_{T_e}\left(\mb{R}\right)   + \delta \hat{\rho}(t),
			\end{equation}
		where $\tau$ is a time over which electronic excitations become decorrelated\cite{Race2010}. 
		Henceforth we neglect the small fluctuating correction $\delta \hat{\rho}(t)$. 

	    The non-conservative force on atom $i$ due to electron lag due to the time-local part is therefore
	        \begin{eqnarray}
	        	\mb{F}_{\mathrm{e-ph}} &\approx& -2 \bf{\tau} \cdot \mathrm{Tr} \left[ \nabla_i \hat{H} \nabla \hat{ \Delta \rho} \right],	\nonumber 	\\
	        	\mbox{so}\quad 
	            \mb{F}_{\mathrm{e-ph},i} &\approx& 2 \tau_i \mathrm{Tr} \left[ \nabla_i \hat{H} \left( \dot{\mb{R}} \cdot \nabla \hat{\rho}_{T_e}  \right) \right],
	        \end{eqnarray}
	    where $\tau_i$ is a typical timescale locally to atom $i$, which we can estimate in terms of the density of states at the Fermi surface $D_i$ as 
	    	\begin{equation}
	    		\tau_i = \zeta \, \frac{\hbar}{k_B T_e} \frac{ D_i }{ \bar{D}}.
	    	\end{equation}
	    $\zeta$ is an empirical constant to be fitted, required by the highly simplified analysis of the electron-phonon coupling given here.
	    $\bar{D}$ is the electron density of states at the Fermi level for the perfect lattice.
	    
	    First-order perturbation theory gives us the gradient of the density matrix from the instantaneous eigenstates
	        \begin{equation}
	            \nabla \hat{\rho}_{T_e} = \sum_{mn} \frac{f(\epsilon_m;\lambda,T_e) - f(\epsilon_n;\lambda,T_e)}{\epsilon_m - \epsilon_n} \left|m\right> \left<m\right| \nabla \hat{H} \left|n\right> \left<n\right|.
	        \end{equation}
	    This RPA-type sum contains the correlated coupling between occupied and unoccupied states which makes electronic susceptibility dependent on features in the electronic density of states.   
	    But our intention is to derive a form suitable for use with interatomic potentials. 
        Typically the model for electronic density of states used with such potentials is very rudimentary, the exception being for bond-order potentials.
        As such it is not reasonable to expect a potential to provide any sensible information beyond a density of states at the Fermi level.        
        We only want to find generic metallic properties near the Fermi surface, and here the sum becomes very much more simple to work with.

\begin{figure}[h!t!b]
           \begin {center}
               \includegraphics [width=5.0in] {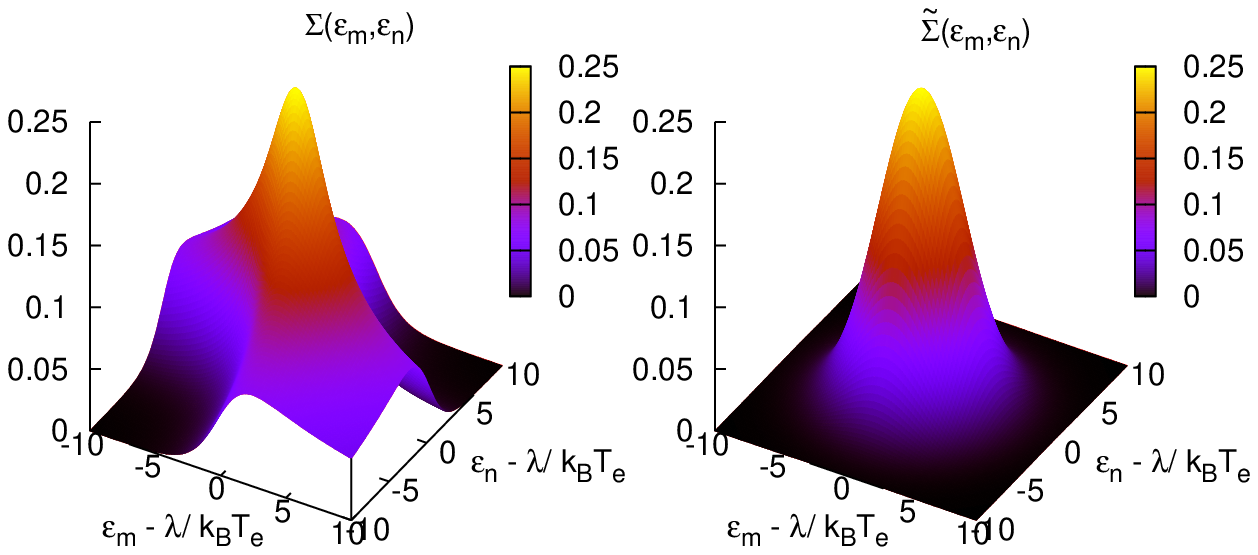}
           \end{center}
           \caption {
           An approximation to the electronic stopping decorrelating occupied and unoccupied states.
            Left: The fully correlated function coupling occupied and unoccupied states shows a peak near the Fermi surface with lobes coupling occupied/unoccupied pairs of states.
            Right: The Gaussian approximation reproduces the peak near the Fermi surface but removes the lobes.
            }
           \label {decorrelation}
       \end {figure}

        Writing $\Sigma = - \frac{f(\epsilon_m;\lambda,T_e) - f(\epsilon_n;\lambda,T_e)}{\epsilon_m - \epsilon_n}$ as the fully correlated coupling function, we can approximate the region near the Fermi surface with a Gaussian product
            \begin{equation}
                \label{decorrelation_eqn}
                \Sigma(\epsilon_m,\epsilon_n) \cong \tilde{\Sigma}(\epsilon_m,\epsilon_n) = \frac{1}{4 k_B T_e} \exp{\left( -\frac{ \left(\epsilon_m - \lambda\right)^2}{2 \sigma^2} \right)} \exp{\left( -\frac{ \left(\epsilon_n - \lambda\right)^2}{2 \sigma^2} \right)}.
            \end{equation}
        Good scaling behaviour can be ensured by adjusting $\sigma$. Equating the expectation value of the two functions on the contour $(\epsilon_m - \lambda)^2 + (\epsilon_n - \lambda)^2 = \sigma^2$ gives $\sigma = 2.8443 k_B T_e$.
        The quality of this approximation is shown graphically in figure \ref{decorrelation}. Note that the approximation is good in the region where an interatomic potential may be sensible, and ignores the contribution from the region where the interatomic potential is likely to be poor.
       The coupling of electronic states separated by $\delta \epsilon = \epsilon_m - \epsilon_n <\approx \sigma$ is well reproduced.
       \footnote{
       It is important to remember that the separation $\delta \epsilon$ is an electronic transition energy and not an ionic kinetic energy- a rule-of-thumb for the conversion can be made as follows.
       We can estimate that an electronic transition $\delta \epsilon$ is stimulated by an ion travelling at speed $v = d \delta \epsilon/h$, where $d\sim \sqrt{3/4}a_0$ is a typical atomic separation and $h$ Planck's constant\cite{Mason_NJP2012}, so at $T_e = 1000K$ the approximation is good for tungsten ions up with kinetic energy up to about 250eV. The range of ion energies handled correctly increases with electronic temperature.
       For keV ions, which might excite larger electronic transitions, the coupling may be incorrectly estimated.  We treat high energy ions in section \ref{estopping}.
       }

        Using this approximation to decorrelate electronic states, the electron lag force is
            \begin{equation}
                \mb{F}_{\mathrm{e-ph},i} \cong - 2 \tau_i \mathrm{Tr} \left[ \left( \dot{\mb{R}} \cdot \left( \frac{\hat{P} \, \nabla \hat{H} \, \hat{P}}{4 k_B T}  \right) \right) \nabla_i \hat{H} \right],
            \end{equation}
        where $\hat{P} = \sum_m \exp{\left( -\frac{ \left(\epsilon_m - \lambda\right)}{2 \sigma^2} \right)} \left| m \right> \left< m \right|$ is a projection operator picking out states near the Fermi surface.
        
        To provide a computationally efficient scheme we need a spatially local functional form for the lag force.
        To do so we can introduce a tight-binding basis of orthonormal atomic orbitals, $\left|i,\alpha\right>$ being the $\alpha$th orbital on the $i$th atom.
        Then we now reject any long-range coupling by approximating $\hat{P} \approx \gamma_i \hat{I}$ in this basis.

        We simplify subsequent calculations by assuming a Friedel model for the density of states; that is a rectangular d-band with width $W_i$ and height $D_i = N_a/W_i$ where $N_a$ is the number of electronic states.
        Then to ensure we retain correct scaling we equate
            \begin{eqnarray}
                \mathrm{Tr}\left[ \hat{P} \right] &=& \gamma_i \mathrm{Tr}\left[ \hat{I} \right] = \gamma_i N_a   \nonumber   \\
                    &=& \int_{-W_i/2}^{W_i/2} \exp{\left( -\frac{ \epsilon^2}{2 \sigma^2} \right)} D(\epsilon) \md \epsilon \nonumber\\
                    &=& \frac{ N_a }{W_i} \sqrt{2 \pi  \sigma^2} \, \mathrm{erf}\left( \frac{ W_i/2} {\sqrt{2} \sigma} \right),    \nonumber   \\
              	\mbox{hence} \quad
           \gamma_i &=& \frac{ \sqrt{2 \pi  \sigma^2}}{W_i}    \, \mathrm{erf}\left( \frac{ W_i/2} {\sqrt{2} \sigma} \right) ,
            \end{eqnarray} 
        which has the limits
            \begin{eqnarray}
                \lim_{T_e \rightarrow 0} \gamma_i &=& \sqrt{2 \pi} \frac{\sigma}{W_i}    \nonumber   \\
                \lim_{T_e \rightarrow \infty} \gamma_i &=& 1.
            \end{eqnarray}
            
        To derive the form for a time- and spatially- local electronic stopping force we introduce an orthonormal set of atomic orbitals where $\left|i_{\alpha}\right>$ is the $\alpha$th orbital on the $i$th atom.
        Then we can introduce $h^{\alpha\beta}(\mb{R}_{ij} ) \equiv \left<i_{\alpha}\right| \hat{H} \left|j_{\beta}\right>$, determining that elements of the electron-ion Hamiltonian operator be a function only of the separation between ions, then
            \begin{eqnarray}             	
            	\label{e-ph_stopping}
                \mb{F}_{\mathrm{e-ph},i} &\cong&
                         - \frac{2 \tau_i }{4 k_B T_e} \sum_{jk,l,\alpha\beta} \left(\gamma_j \gamma_k \dot{\mb{R}}_l \cdot \left<j_{\alpha}\right|  \nabla_l \hat{H} \left|k_{\beta}\right> \right) \left<k_{\beta}\right| \nabla_i \hat{H} \left|j_{\alpha}\right> \nonumber \\
                    &=&
                         \frac{\tau_i }{k_B T_e} \sum_{j,\alpha\beta} {\left( \gamma_i \gamma_j ( \dot{\mb{R}}_j - \dot{\mb{R}}_i ) \cdot \nabla_i h^{\alpha\beta}(\mb{R}_{ij}) \right)} \, \nabla_i h^{\alpha\beta}(\mb{R}_{ij}) \nonumber \\
                    &=& \zeta  \frac{ \hbar \bar{W} }{W_i (k_B T_e)^2} \,  \sum_{j \in \mathcal{N}_i ,\alpha\beta} \, \frac{2 \pi  \sigma^2}{W_i W_j} \left( \mathrm{erf}\left( \frac{ W_i/2} {\sqrt{2} \sigma} \right)   \mathrm{erf}\left( \frac{ W_j/2} {\sqrt{2} \sigma} \right)    \,  \nabla_i h^{\alpha\beta}(\mb{R}_{ij}) \cdot ( \dot{\mb{R}}_j-\dot{\mb{R}}_i  ) \right) \,  \nabla_i h^{\alpha\beta}(\mb{R}_{ij}) \nonumber   \\
                    &\equiv& \frac{  \bar{W} }{W_i} \sum_{j \in \mathcal{N}_i} \, \mb{B}_{ij} ( \dot{\mb{R}}_j-\dot{\mb{R}}_i )
            \end{eqnarray}
        defining the environmentally dependent damping tensor $\mathbf{B}_{ij}$.
        
        At low temperatures this tends to 
            \begin{equation}
                \lim_{k_B T_e \ll W_i} \mb{B}_{ij} =  \zeta   \frac{ 50.831 \hbar  }{W_i W_j} \sum_{\alpha\beta} \nabla_i h^{\alpha\beta}(\mb{R}_{ij}) \nabla_i h^{\alpha\beta}(\mb{R}_{ij})^T,
            \end{equation}
        and at high temperatures
        	\begin{equation}
        		\lim_{k_B T_e \gg W_i} \mb{B}_{ij} = \zeta  \frac{ \hbar}{(k_B T_e)^2}  \sum_{\alpha\beta} \nabla_i h^{\alpha\beta}(\mb{R}_{ij}) \nabla_i h^{\alpha\beta}(\mb{R}_{ij})^T,
        	\end{equation}
        thus having the correct form for both angular and temperature dependence for a finite band model\cite{Mason2007}.      
        In section \ref{estopping} we revisit the high temperature dependence where the assumption of a finite band model breaks down.
        This will also tackle the case where a swift ion is able to excite electrons over an energy comparable to the band width.

        Note that this is the low-ion-energy form for the electron-phonon damping tensor, and should be applied to all metal ions regardless of kinetic energy.
        
    \subsection{Empirical potentials}        
        
		We must now relate elements of the Hamiltonian operator $h(\mb{R}_{ij} )$ and the width of the d-band $W_i$ to an empirical potential.
		The embedded atom potential form expresses the energy of the system of atoms as	
			\begin{equation}
				E\left( \mb{R} ; t \right)_{\mathrm{EAM}} = \half \sum_i M_i \left| \dot{\mb{R}}_i \right|^2 + V\left( \mb{R} \right) + \sum_i F\left[ \rho_i \right],
			\end{equation}
		where $\rho_i = \sum_{j \in \mathcal{N}_i} \phi( \mb{R}_{ij})$ is a scalar proportional to the electron density into which atom $i$ is embedded.
		For the original Finnis-Sinclair form\cite{Finnis1984} the embedding function is a simple square root function $F\left[ \rho \right] = -A \sqrt{ \rho }$.
		If we assume a Freidel rectangular d-band model with $N_e$ electrons in $N_a$ states, width $W_i$ and height $D_i = N_a/W_i$ with centre of band $\alpha_i$,
		the zero-temperature bond energy is given by
			\begin{equation}
				F_{T=0,i} = -A \sqrt{ \rho_i } = \int_{\alpha_i-W_i/2}^{\lambda} 2 D_i(\epsilon) (\epsilon-\alpha_i)  \md \epsilon = - \frac{N_e W_i}{2} + \frac{N_e^2 W_i}{4 N_a} 
			\end{equation}
		hence the width of the d-band is given by 
			\begin{equation}
				W(\rho) = \frac{ 4 A N_a }{ N_e (2 N_a - N_e ) } \sqrt{\rho} \equiv w \sqrt{\rho}, 
			\end{equation}
		defining the constant $w$.
    	Note that for the DND potential $w=\bar{W}$ as these potentials are normalised to $\rho=1$ at the equilibrium lattice spacing.
		The second moment of the density of states is\cite{FinnisBook}
			\begin{equation}
				m^{(2)}_i \equiv \sum_{j \in \mathcal{N}_i,\alpha\beta} \left( h^{\alpha\beta}( \mb{R}_{ij}) \right)^2 = \int_{\alpha_i-W_i/2}^{\alpha_i+W_i/2} 2 D_i(\epsilon) (\epsilon-\alpha_i)^2 \md \epsilon  = \frac{ N_a W_i^2 }{6}.
			\end{equation}
        We now must flatten out the angular orbital dependence and generate a functional form for the damping dependent only on the distance between atoms. 
        This can be done by equating
            \begin{equation}
                \left( h^{\alpha\beta}( \mb{R}_{ij}) \right)^2 = \frac{N_a}{6} w^2 \left( \phi( \mb{R}_{ij}) \right)^2.	
            \end{equation}

        At low temperatures the damping tensor is therefore 
            \begin{equation}
                \lim_{k_B T_e \ll W_i} \mb{B}_{ij} =  \zeta \frac{ 8.4719 N_a w^2 \hbar  }{W_i W_j} \nabla_i \phi(\mb{R}_{ij}) \nabla_i \phi(\mb{R}_{ij})^T.
            \end{equation}

	\section{Electronic Entropy}
	    \label{electronicEntropy}
		The electronic entropy for atom $i$ is given by integrating over the Fermi-Dirac occupations:
            \begin{eqnarray}
                S_{e,i} &=& - 2 k_B \frac{N_a}{W_i} \int_{\alpha_i - ^{W_i}\!\!/\!_2}^{\infty} f \log f + (1-f) \log (1-f) \, \md \epsilon \nonumber   \\
                    &=& 4 k_B^2 T_{e} \frac{N_a}{W_i} \left[ \frac{\pi^2}{6} +\frac{\mu_i^2}{2} + \mathrm{Li}_2 \left( -e^{-\mu_i} \right)  \right] - k_B N_e \mu_i.
            \end{eqnarray}
        As $T_e/W_i \rightarrow 0$, this expression recovers the Sommerfeld limit for the free electron gas:
            \begin{equation}
                \lim_{T_e/W_i \rightarrow 0} S_{e,i} = \frac{ \pi^2}{3} \, k_B^2 T_e \, \left(2 \frac{N_a}{W_i}\right),
            \end{equation}
        and for $W_i/T_e \rightarrow 0$ is bounded
            \begin{equation}
                \lim_{W_i/T_e \rightarrow 0} S_{e,i} = 4 k_B \frac{N_a}{W'} \left[ \frac{\pi^2}{6} +\frac{\mu_0^2}{2} + \mathrm{Li}_2 \left( -e^{-\mu_0} \right)  \right] - k_B N_e \mu_0.
            \end{equation}          
            
       	We can use this to construct an electronic free energy, but first we must consider when to use it.
        The vast majority of MD simulations are assumed to be performed in the zero electronic temperature Born-Oppenheimer limit.
        Here we offer a choice of microcanonical or canonical embedding functions, which should match the choice of electronic heat transfer model. 

        We might assume the electrons are instantaneously in equilibrium with a thermal reservoir and so at a constant temperature.
		This is the case where there is no electronic stopping, and so may be suitable for finite temperature simulations where ions and electrons are in thermal equilibrium.
		Work must be done by, or against, the heat bath to maintain the temperature of the electrons.
		Then the correct embedding function is $F_{T_e,i} = F_{0,i} + \Theta_{ T_{e},i } - T_{e,i} S_{e,i}$.
		
		The alternate limit is to consider the electron evolution isentropic, so the ion-electron system is microcanonical. 
		This is the case if electronic stopping is used, as we assume work is done in raising an electron pseudotemperature, which then subsequently thermalises by electron-electron collisions.
		Then the correct embedding function is $F_{T_e,i} = F_{0,i} + \Theta_{ T_{e},i }$.

		The consequence of this decision may be important if high temperatures are reached.
		If the evolution is microcanonical, the effect of raising electronic pseudotemperature is to stiffen the bonds.
		If the evolution is canonical, a finite electronic reservoir temperature has the effect of relaxing the bonds.
		
		We can now fix the bandwidth at low density:
			In the microcanonical ensemble 
				\begin{equation}
					\lim_{W_i/T_e \rightarrow 0} F_{T_e,i} = - \frac{ N_e W' }{2}(k_B T_{e}) + \left(\frac{2 N_a}{W'}\right) (k_B T_{e}) \left[ \frac{\pi^2}{6} + \frac{\mu_0^2}{2} + \mathrm{Li}_2\left( -e^{-\mu_0} \right)  \right].
				\end{equation}
		As this expression is linear in temperature, we can insist that isolated atoms have zero energy, and so also have zero electronic heat capacity, by solving $F_{T_e,i} =0 $ for $W'$ in this limit.
		Values for $W'$ are given in table \ref{parameterization}.

\section{Fermi Velocity}
    \label{FermiVelocityCalc}
			
		The average Fermi velocity $v_F$ can be computed for transition metals using density functional theory.
		\begin{equation}
			\label{FermiVelocity}
			v_F = 	\sum_l \oint \left| \frac{\partial E_{\mb{k},l}}{\partial \mb{k}} \cdot \mb{n} \right| \md S,
		\end{equation}
		where $l$ is a band index and the integral is over the Fermi surface.
		We used the abinit code\cite{Abinit}, using the PAW method\cite{Torrent_CMS2008} and a GGA-PBE functional\cite{Perdew_PRL1996} to construct the Kohn-Sham eigenstates for V,Nb,Ta,Cr,Mo and Fe.
		Spin degeneracy was lifted for the calculation in iron.
		The spin-orbit interaction was included for tungsten, using the HGH semicore pseudopotential\cite{Hartwigsen_PRB1998} and LDA functional\cite{Perdew_PRB1981}.
		To converge the Fermi velocity it is necessary to use a very highly refined Fermi surface.
		This we acheived by interpolating eigenvalues at $32^3$ $k$-points for each band in turn to a piecewise cubic spline, and then finding the surface on a grid of $128^3$ interpolated $k$-points using the marching cubes method.
		The results are in table \ref{thermalConductivityT}.
		The density of states was also found from this high-resolution triangulation of the Fermi surface using the divergence theorem: the volume encapsulated by a closed surface of triangles for which the $i$th triangle has vertices $\mb{x}_{ij}$ is given by 
			\begin{equation}
				V = \sum_i  \frac{1}{6} \frac{ \mb{n}_i \cdot \sum_j \mb{x}_{ij}  }{ \left|\mb{n}_i\right| }
			\end{equation}
		where $\mb{n}_i = ( \mb{x}_{i2}-\mb{x}_{i1} ) \times ( \mb{x}_{i3}-\mb{x}_{i2} )$ is the signed unnormalised normal for triangle $i$.
		The density of states is then 
			\begin{equation}
				D(\epsilon_F) \equiv 2 \sum_l \frac{1}{V_{BZ}} \frac{\partial V_l}{\partial \epsilon},
			\end{equation}
		where $V_{BZ}$ is the volume of the Brillouin zone, $l$ is a band index for those bands crossing the Fermi level and the 2 is for spin degeneracy.
       \begin{figure}[h!t!b]
           \begin {center}
               \includegraphics [width=5.0in] {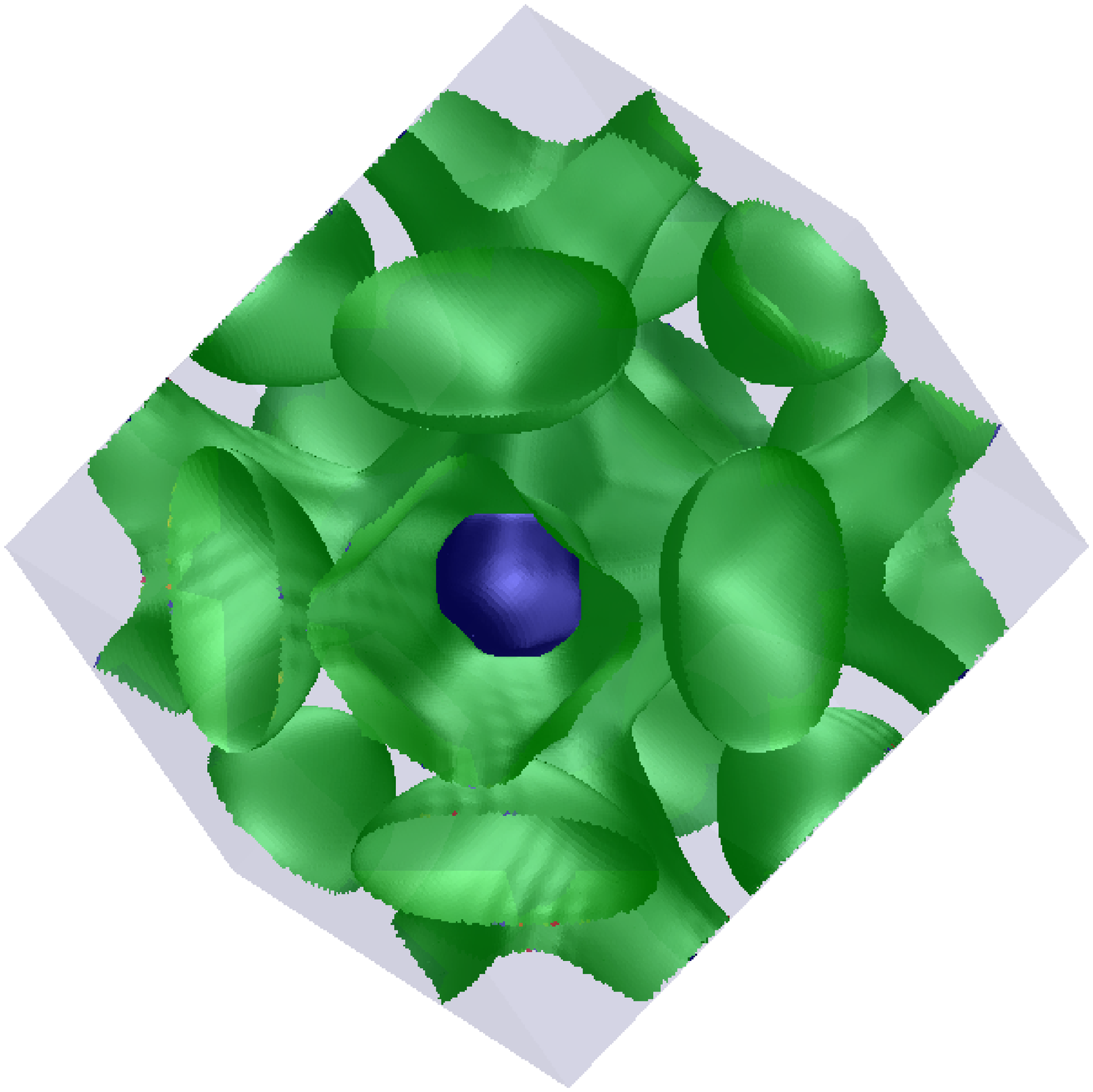}
           \end{center}
           \caption {
           Fermi surface of tantalum constructed from a triangulation of a cubic-spline interpolation of $32^3$ $k$-points.
            }
           \label{taFermiSurface}
       \end {figure}
       Figure \ref{taFermiSurface} shows the triangulated Fermi surface of tantalum.
		\begin{table}
	        \centering
	        \small
	        \begin{tabular}{l|lll}
	        	Element &	$v_F$ 	  	& 	$D(E_F)$		& $\lim_{T_e\rightarrow 0}$ $C_e/T_e$  \\
	        			&	(\AA/fs)	&	(1/eV)  		& (x$10^{-9}$ eV/K$^2$/$\AA^3$)       \\
	        	\hline                                                			
	        	V       &	4.47	  	&	1.81 			& 3.148 	                          \\
	        	Nb      &	6.46	  	&	1.49 			& 2.024 	                          \\
	        	Ta      &	6.47	  	&	1.31 			& 1.772 	                          \\
	        	Cr      &	5.73	  	&	0.683			& 1.390 	                          \\
	        	Mo      &	8.72	  	&	0.600			& 0.940		                          \\
	        	W       &	9.50	  	&	0.355			& 0.546		                          \\
	        	Fe		&	4.95	  	&	1.01 			& 2.095		                          \\
	        \end{tabular}                                                                                                                                                                                                        
	        \caption{                
	        	The Fermi velocity, density of states at the Fermi level and low temperature heat capacity coefficient calculated using DFT.
	        }
	        \label{thermalConductivityT}
	    \end{table}				
		
		The Fermi velocity is expected to be a function of strain, as for the free electron gas
			\begin{equation}
				v_F = \frac{\hbar}{m} \left( \frac{ 3 \pi^2 N_e }{\Omega} \right)^{1/3} \sim \frac{1}{\sqrt{D(\epsilon_F)}}
			\end{equation}
		In an MD simulation estimating the atomic volume $\Omega$ is not easy only-the-fly, as it relies on subdividing the simulation cell into regions associated with each atom- a procedure ill-defined at surfaces.		
		For this particular implementation of thermal properties based on the Finnis-Sinclair potential form we can propose using the bandwidth as a proxy for the inverse length-scale, so that the Fermi velocity to use for atom $i$ is 
			\begin{equation}
				\label{localFermiVelocity}
				v_{F,i} = v_F \sqrt{ \frac{ W_i }{ \bar{W} } }.
			\end{equation}

\end{document}